\documentclass{aa}

\usepackage{natbib, twoopt}
\usepackage{newtxtext,newtxmath}
\usepackage[T1]{fontenc}
\usepackage[normalem]{ulem}
\usepackage{acronym}

\usepackage{graphicx}   % Including figure files
\usepackage{amsmath}    % Advanced maths commands
\usepackage{amssymb}    % Extra maths symbols
\usepackage[usenames,dvipsnames]{xcolor}
\usepackage{soul}
\usepackage{enumitem} 
\usepackage[breaklinks=true]{hyperref}
\usepackage{xcolor} % For colors
\usepackage[normalem]{ulem} %For striking out text
\usepackage{todonotes} % for the color boxy things
\usepackage{makecell}

\bibpunct{(}{)}{;}{a}{}{,}

\newcommand{\KT}[1]{\textcolor{Periwinkle}{ #1}}

\newcommand{\kms}{\,{\rm km\,s^{-1}}}   % per cm-squared
\newcommand{\msun}{\,{\rm M_\odot}}     % per cm-squared
\newcommand{\kpc}{\,{\rm kpc}}  % per cm-squared
\newcommand{\Gyr}{\,{\rm Gyr}}  % per cm-squared
\newcommand{\pc}{\,{\rm pc}}    % per cm-squared
   
\newcommand{\Myr}{\,{\rm Myr}}  % per cm-squared

\newcommand{\unitbubblerate}{\,{\rm kpc^{-2}~Gyr^{-1}}}
\newcommand{\accunit}{\,{\rm km^2\,s^{-2}\,kpc^{-1}}}
\newcommand{\pd}{\partial} 
\newcommand{\Sgas}{\Sigma_{\rm gas}}
\newcommand{\vr}{v_{\rm R}}
\begin{document}

\title{Superbubbles as the source of dynamical friction: Gas migration, and stellar and dark matter contributions}
\author{
Rain Kipper\inst{1}\thanks{E-mail: rain.kipper@ut.ee} \and
Indrek Vurm\inst{1}\and
Aikaterini Niovi Triantafyllaki\inst{1}\and
Peeter Tenjes\inst{1}\and
Elmo Tempel\inst{1,2}
}
\institute{Tartu Observatory, University of Tartu, Observatooriumi 1, T\~oravere 61602, Estonia
\and Estonian Academy of Sciences, Kohtu 6, 10130 Tallinn, Estonia
}
\date{Received 30/04/2024 / Accepted 31/10/2024}

\label{firstpage}
\abstract{
The gas distribution in galaxies is smooth on large scales, but is usually time-dependent and inhomogeneous on smaller scales. The time-dependence originates from processes such as cloud formation, their collisions, and supernovae (SNe) explosions, which also create inhomogeneities. The inhomogeneities in the matter distribution give rise to variations in the local galactic gravitational potential, which can contribute to  dynamically coupling the gas disc to the stellar and the dark matter (DM) components of the galaxy. Specifically, multiple SNe occurring in young stellar clusters give rise to superbubbles (SBs), which modify the local acceleration field and alter the energy and momentum of stars or DM particles  traversing them, in broad analogy to the dynamical friction caused by a massive object. Our aim is to quantify how the acceleration field from SBs causes dynamical friction and contributes to the secular evolution of galaxies. In order to assess this, we constructed the time-dependent density modifications to the gas distribution that mimics a SB. By evaluating the acceleration field from these density modifications, we were able to see how the momentum or angular momentum of the gas hosting the SBs changes when stars pass through the SB. Combining the effects of all the stars and SBs, we constructed an empirical approximation formula for the momentum loss in homogeneous and isotropic cases. We find that the rate at which the gas disc loses its specific angular momentum via the above process is up to 4\% per Gyr, which translates to under one-half of its original value over the lifetime of the disc. For comparison, the mass transfer rate from SBs is about one order of magnitude less than from gas turbulence, and hence the SB contribution should be included to account for the gas migration rate more accurately than $10\%$. Finally, we studied how the dynamical coupling of the gas disc with the DM halo depends on assumptions on the halo kinematics (e.g. rotation) and found a $\sim 0.3$~\% variation in the gas disc secular evolution between different DM kinematic models.
}

\maketitle

\section{Introduction}
\label{sec:introduction}
In this paper our aim is to study two aspects of galaxy structure. These are the gas inhomogeneities that cause dynamical friction in gaseous discs of galaxies and the corresponding secular evolution of these gas discs. Specifically, in this paper we focus on superbubbles (SBs) created by supernova (SN) explosions in gas discs  as the source of dynamical friction. 

\begin{figure}
    \centering
    \includegraphics{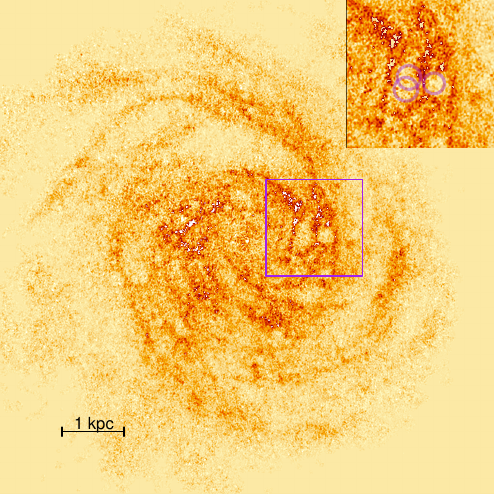}
    \caption{Illustration of SBs in galaxy NGC 6946 from the intensity map of the THINGS survey \citep{Walter:2008}. The inset in the top right corner is a zoomed-in  image of a fraction of the galaxy where the locations of three SBs are shown (pink circles). A more detailed map of all the bubbles in the galaxy is shown in \citet{Brinks:2007}. }
    \label{fig:ngc6946}
\end{figure}

Superbubbles  are cavities in the gas distribution;  they span from hundreds to thousands of parsecs. \citet{Nath:2020} compared the size distributions of observed HI holes from THINGS \citep{Walter:2008} with those derived from detailed hydrodynamical simulations, and concluded that SBs are caused by multiple SNe in a large OB association or in multiple associations. Many observations document the phenomenon as dark regions in the gas discs of galaxies (e.g.  \citealt{Lara-lopez:2023}). Recent studies using JWST observations give excellent samples of SBs in NGC~628 \citep{Watkins:2023}. In addition, studies such as \citet{Brinks:2007, Walter:2008} provide good observations of holes in the HI distribution in NGC~6946  (see Fig.~\ref{fig:ngc6946}). The association of HI holes and SN explosions has been studied in detail (e.g. \citealt{Lara-lopez:2023, Sarbadhicary:2023, Barnes:2023}).

From a physical perspective,
SBs are  underdense regions within the gaseous medium, filled by hot pressurised gas. This is  created by multiple SN explosions that inflate a cavity within the ambient medium, driving a shock wave into the latter and accumulating gas at its edges \citep{Low:1988}. When the SN energy input ceases, the SB starts to dissolve. The hot interior exchanges the heat with edges via thermal conduction \citep[]{Yadav:2017} (hereafter Y17),
causing heat to diffuse towards the cold shell and the shell material to evaporate towards the interior of the bubble \citep{Cowie:1977}. During the two phases, the shell of the SB is either expanding or shrinking, causing disturbances in density and gravitational potential. An overview of these processes is provided by \citet{Keller2017SuperbubbleFI}. Physically, the situation is rather complex, as the heat conduction may depend on magnetic fields, and the initial gas distribution may be clumpy instead of  homogeneous. These aspects may be relevant and were studied in \citet{Keller:2014}. 

Our aim was to expand the study of SBs and their gravitational interaction with the galaxy disc, as the current status of this research is not extensive. \citet{GarciaConde:2023}, while not aiming to solve the SBs influence to the structure of galaxies, notes that in  low-density environments, and in connection to bending waves,  SNe can redistribute gas and cause acceleration field changes. In the context of dwarf and high-redshift clumpy galaxies, the strong scattering of giant clouds and holes may help to form exponential-like discs \citep{Struck_2017, Wu_2020}. However, these works concentrated on turbulent rather than quiet disc galaxies.
 
As a rule, the gas disc and the stellar disc of a galaxy rotate at rather different velocities: the stellar disc has a slower rotation.  Therefore, there is, on average, a systematic mismatch between the velocities of the stars and the SBs, which can give rise to a net momentum and/or energy transfer between the stars traversing the SB and the gaseous environment. In the case of a massive perturber in a stellar environment, this is called dynamical friction \citep{Chandrasekhar:1943}. In \citet{Kipper:2020, Kipper:2023} we developed a method to calculate dynamical friction effects in a non-homogeneous gravitational potential. For the present paper we used the method for the case of SBs. For this we needed to know the  density contrast (positive or negative) of a SB relative to the host environment. 

Dynamical friction acts in two ways: it changes  the velocities of stars and the velocities of SBs, both of which result in the evolution of the disc. 
These changes are slow (secular). There are various other processes causing the secular evolution of galaxies \citep{Kuzmin:1961, Kuzmin:1963, Binney:2013, Kipper:2021, Husko:2023} (also see the review by \citet{Kormendy:2013}). Getting a grasp on these processes is necessary in order to  reach a level of  understanding   to reproduce all galaxy aspects in simulations. A good understanding of the baryonic effects also allows  dark matter (DM) studies to be improved \citep{Maxwell:2015, 2020PhRvD.101j3023B}. The list of internal processes that contribute to secular evolution includes the scattering of stars,  characterised by the Fokker-Planck equation (see \citealt{Binney:2008}); churning, which is the change in angular momentum by spiral, bar, or other non-axisymmetry \citep{Sellwood:2002, Minchev:2010, Vera_Ciro_2014, Patil:2023}; blurring, where the breadth of an orbit grows \citep{Sch:2009} out to very large galactic radii \citep{Lian:2022}; and local disc instabilities \citep{Husko:2023}. These processes provide a significant redistribution of matter in galaxies. In the present paper we study an addition to these effects that might be the reason behind a slow redistribution of stars and gas. Similarly to giant molecular clouds acting as scatterers of stars, we investigate how much the SBs alter the gas distribution. 

The purpose of the current paper is to investigate whether the time-dependent gravitational potential disturbances help the transfer of angular momentum between the gas disc and stellar disc in a smooth and relatively unperturbed (regular) galaxy. Thereafter, we consider whether  these contributions are sufficient to cause a significant migration of the gas disc.

In Sect. \ref{sec:SBdesc} we characterise our description of the SB and its evolution. In Sect.~\ref{sec:analytical_SB} we show how passage  through the SB influences the single-star kinematics numerically, and we provide analytical approximations for it. In Sect.~\ref{sec:exmamples} we show numerically the transfer of momentum or angular momentum in various environments. We discuss the results in Sect.~\ref{sec:discussion}, and summarise in Sect.~\ref{sec:summary}. 

\section{Description of a superbubble}\label{app:bubble_calc_formulae}
\label{sec:SBdesc}
Although SBs may have a complex morphology \citep{Kim:2017},  in essence they are underdense regions embedded within a background gaseous environment (the host). In our present treatment we consider this background to be homogeneous, and describe the SB as a spherical region with (formally) negative interior density relative to the background, surrounded by a slightly overdense shell of material displaced from the interior. We  subsequently consider the effect of this configuration on the local acceleration field relative to that of the unperturbed host.

\subsection{The density profile}\label{sec:density_profile}
\begin{figure}
    \centering
    \includegraphics{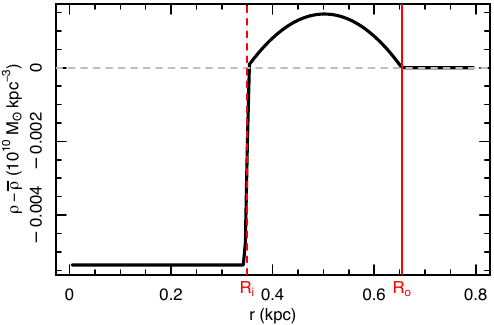}
    \caption{Density contrast profile approximation of the SB at $t = 30\Myr$ inspired by the simulated SB behaviour in Y17 and described in Sect.~\ref{sec:density_profile}. }
    \label{fig:bubble_density_profile}
\end{figure}

We approximated the SB density profile at any moment in time using two parameters: its inner and outer radius, $R_{\rm i}$ and $R_{\rm o}$, respectively. The inner radius is the radius closer to the centre of the cavity before the overdensity region, and the outer radius the one closer to the interstellar medium (ISM) at  the edges of the SB. For convenience, we also denote their midpoint $R_{\rm m} = (R_{\rm i} + R_{\rm o})/2$. The approximation of the SB   mimics the profile from Y17, where they simulated the evolution of a SB in a homogeneous gas environment with initial density $\rho_{\rm env} = 0.015\,{\rm M_\odot pc^{-3}}$. The profile itself is shown in Fig. A1 of Y17. Our approximation assumes that the SNe have pushed out all the gas inside $R_{\rm i}$. Hence, the physical density inside it is zero, but in the form of density-correction $\rho^*\equiv \rho - \rho_{\rm env}$ its value is $-\rho_{\rm env}$. In our description the gas from inside is pushed between $R_{\rm i}$ and $R_{\rm o}$, where we describe the density distribution as a quadratic polynomial. Outside of $R_{\rm o}$, the SB has had no influence and the density-correction is zero. Combining all the regions together, we have the formula
\begin{equation}
\rho^* \equiv \rho - \rho_{\rm env} = \left\{ \begin{aligned} 
  & - \rho_{\rm env}, & R < R_{\rm i}\\
  & B - A(R - R_{\rm m})^2,& R_{\rm i} \le R \le R_{\rm o}\\ \label{eq:bubble_density_profile}
  & 0, & R > R_{\rm o}
\end{aligned} \right.
.\end{equation} %eq:bubble_density_profile
The coefficients $A$ and $B$ are found from mass conservation: 
the mass shifted from within radius $R_{\rm i}$ equals that added between $R_{\rm i}$ and $R_{\rm o}$ and at $\rho^*(R_i) = \rho^*(R_o) = 0$. Figure~\ref{fig:bubble_density_profile} shows this profile visually. 
%fig:bubble_density_profile

\subsection{The evolution of  superbubbles}\label{sec:bubble_evolution}
%The second aspect of SB approximation is its evolution. 
Superbubbles are short-lived compared to galaxies;  they vanish soon after the SNe of the star-forming region cease. We split the description of the evolution into two phases: the expanding phase, where the SNe are pushing the gas out, and the vanishing phase, where the SNe support ceases, and the SBs start to shrink. We call this phase the collapsing phase to emphasise our assumption. We denote the time of the split between the phases as $t_{\rm max}$. 

The precise evolution of  the SBs is uncertain. The formation of the SBs is studied in many magnetohydrodynamic simulations \citep{Kim17TIGRESS, Girichidis:2016}, but their complete evolution depends not only on the SN explosions, but also on the physics responsible for the collapse of the SB cavity. In the literature, the first stages of SB evolution have been worked through thoroughly. In the case of a single SN explosion, the main physical processes have been analysed (see e.g.  \citealt{Ryden:2021}), but in the case where many SNe explode in spatial and temporal vicinity, it becomes more complicated, and these solutions have  only been simulated (e.g. \citet{Kim:2017}). In the collapsing phase, turbulences and shearing motions redistribute former shell material back into the SB volume \citep{Kim17TIGRESS}. Turbulences are chaotic phenomena that do not follow strict descriptions, which makes analytical approximation somewhat difficult. From the analytical side,  the collapsing stage in atmospheric conditions   was studied by   \citet{Penner:2003}, among others,  who showed that after the shockwave travels through the atmosphere, the outer edge of the shell stalls and inner edge turns into a cooling wave that moves the interiors of the cavity. However, this does not necessarily align with what is happening in the case of SBs, as the cooling is different for atmospheric chemistry and the ISM.  

\begin{figure}
    \centering
    \includegraphics{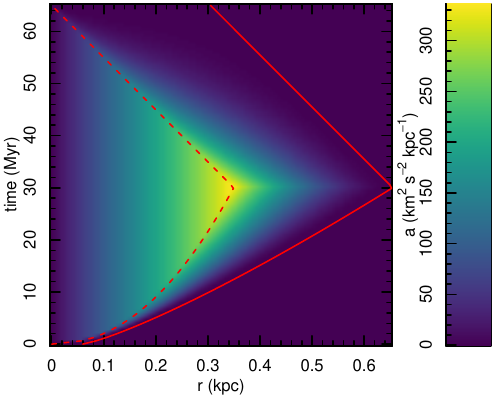}
    \caption{Acceleration field of SB. The solid and dashed red lines are    the outer and inner radius of the fiducial model from Y17, and  are described with Eqs.~\eqref{eq:Ri} and \eqref{eq:Ro}. The colour scale shows the radial acceleration values described in Sect.~\ref{sec:analytical_SB}. }
    \label{fig:accfield_of_bubble}
\end{figure}

The Y17 simulation describes only the first $30\,\Myr$ of the SB evolution as this is the approximate age when the least massive core-collapse SNe explode. Thus, we denote this as the expansion phase. We approximated  the evolution of both $R_{\rm i}$ and $R_{\rm o}$ as a power law with two and three free parameters, respectively. For the vanishing of the SB phase, which was not covered by the Y17 simulation, we
assumed a scenario describable with a characteristic speed, which  we treated as constant. It mimics a collapse   occurring for both $R_{\rm i}$ and $R_{\rm o}$ with a constant speed $v_{\rm col}$. We denote the lifetime of the SB as $t_{\rm end}$,   defined as the time when the $R_{\rm i}$ shrinks back to $0\kpc$. The analytical form for $R_{\rm i}$ is
\begin{equation}
    R_{\rm i}^{\rm fid} = \left\{ \begin{aligned} 
      &  R_{\rm u}(t / t_{\rm i,e} )^{\alpha_{\rm i, e}}, & t \le t_{\rm max}\\
      & R_{\rm max,i} - v_{\rm col}(t-t_{\rm max,i})& t_{\rm max} < t \le t_{\rm end}\\
      & 0, & t > t_{\rm end}
    \end{aligned} \right.\label{eq:Ri_fid}
\end{equation} %eq:Ri_fid
and for $R_{\rm o}$
\begin{equation}
    R_{\rm o}^{\rm fid} = \left\{ \begin{aligned} 
      &  R_{\rm o0} + R_{\rm u}(t / t_{\rm o, s} )^{\alpha_{\rm o}}, & t \le t_{\rm max}\\
      & R_{\rm max,o} - v_{\rm col}(t-t_{\rm max,i})& t > t_{\rm max}
    \end{aligned} \right..\label{eq:Ro_fid}
\end{equation} %eq:Ro_fid
We approximated the parameters in this formula to match the evolution provided in  Y17, and derived the following values: 
$t_{\rm i,e} = 290\Myr$, $\alpha_{\rm i, e} = 0.47$, $R_{\rm o0} = 60\pc$, $t_{\rm o, s} = 56.7\Myr$, $\alpha_{\rm o} = 0.815$. In addition, we   denoted $R_{\rm max, i}$ and $R_{\rm max, o}$ as the SB inner and outer size at $t_{\rm max}$; $v_{\rm col}$ as the collapsing speed; and the superscript `fid' as the fiducial run of the simulation. $R_{\rm u}$ is a constant with unit-length for dimensional considerations. The evolution of $R_{\rm i}^{\rm fid}$ and $R_{\rm o}^{\rm fid}$ is shown in Fig.~\ref{fig:accfield_of_bubble} as dashed and solid red lines, respectively. 

Our currently used approximation of the collapsing phase is when the inner and outer parts of the SB collapse with the same speed. This is not necessarily the case. In order to see how different assumptions change, we analysed our results (see  Appendix~\ref{app:SB:timedep}) and we tested the cases where (i) $R_{\rm i}$ and $R_{\rm o}$ collapse at different speeds; (ii) the SB does not collapse, but mixes with turbulence characterised by modulation of the amplitude of acceleration; and (iii) SB disappears due to shearing motions caused by differential rotation of the galaxy.

Although   Y17    concentrates on their fiducial model, they provided some scaling relations for different SBs. In the current case, the relevant relation is the scaling of the SB radii depending on the number of progenitor SNe. The scaling behaves as the modifier of radii:  $R_{\rm i}$ and $R_{\rm o}$ are found from fiducial values by multiplying them with the term $s_{r}$, or equivalently 
\begin{equation}
s_{\rm r} = \frac{R_{\rm i}}{R_{\rm i}^{\rm fid}} = \frac{R_{\rm o}}{R_{\rm o}^{\rm fid}}.\label{eq:sr_def}
\end{equation} % eq:sr_def
Y17 shows that $s_{\rm r}$ depends on the number progenitor SNe as $N_{\rm SN}^{1/5}$. The fiducial model corresponded to $100$ SNe, and hence 
\begin{eqnarray}
    s_{\rm r} = \left(\frac{N_{\rm SN}}{100}\right)^{1/5}. \label{eq:sr_fiductial}
\end{eqnarray}%eq:sr_fiductial

To match the number of SNe with the mass of their progenitor open clusters we use the initial mass function (IMF) $\Psi\,{\rm d}m$. We match the IMF normalisation to the mass of the SB progenitor open cluster ($M_{\rm OC}$) by introducing a normalising constant $n_{\rm IMF}$ so that 
\begin{eqnarray}
    \int\limits_0^{\infty}mn_{\rm IMF}\Psi(m){\rm d}m = M_{\rm OC}. \label{eq:IMF_normalisation}
\end{eqnarray} %eq:IMF_normalisation

Secondly, we are interested in the number of stars that can produce core-collapse SNe. In terms of IMF it is an integral between the suitable mass range ($m_{\rm SN,min}\dots m_{\rm SN, max}$):
\begin{equation}
    N_{\rm SN} = \int\limits_{m_{\rm SN,min}}^{m_{\rm SN,max}}n_{\rm IMF}\Psi(m)\,{\rm d}m = 
    M_{\rm OC}\frac{\int\limits_{m_{\rm SN,min}}^{m_{\rm SN,max}}\Psi(m)\,{\rm d}m}{\int\limits_0^{\infty}m\Psi(m)\,{\rm d}m}.
\end{equation} %IMF -> N_OB
The division of integrals provides the number of SNe per unit cluster mass, which we denote as $\vartheta$ so that $N_{\rm SN} = \vartheta M_{\rm OC}$. 

For the numerical evaluation of $\vartheta$ we need to specify $\Psi$, $m_{\rm SN,min}$, and $m_{\rm SN,max}$. For the $\Psi$ we selected the Kroupa IMF \citep{Kroupa:2001}, the $m_{\rm SN,min} = 8\msun$, but the $m_{\rm SN,max}$ is not well determined. There have been  dimming periods observed  of massive stars, and hints of faint or absent SNe with masses above $m_{\rm SN, max} = 17\msun$ \citep{Reynolds:2015, Byrne_2022}, which shows that there is a lack of SNe with masses above $m_{\rm SN, max} = 17\msun$, although this has  not yet been confirmed. 
Therefore, it is possible that the highest mass stars do not explode as SNe and we should adopt $m_{\rm SN, max} = 17\msun$, but it is unclear if this affects all  massive stars or only some fraction of them. The conventional text-book value is still $m_{\rm SN, max} = \infty\msun$ for integrations. In these cases, the values are $\vartheta_{17} = 0.0054\,{\rm M_\odot^{-1}}$ and $\vartheta_{\infty} = 0.0084\,{\rm M_\odot^{-1}}$. In the current paper we adopt the more classical $\vartheta = \vartheta_{\infty}$. 
%0.00843961967

Combining all of this together, we see that for the cluster mass of $M_{\rm OC}$ the inner and outer radii are determined from
\begin{eqnarray}
    R_{\rm i} &=& R_{\rm i}^{\rm fid}(\vartheta M_{\rm OC}/100)^{1/5},\label{eq:Ri}\\
    R_{\rm o} &=& R_{\rm o}^{\rm fid}(\vartheta M_{\rm OC}/100)^{1/5}.\label{eq:Ro}
\end{eqnarray} %Ri, Ro from fiducial

\subsection{Selecting open clusters able to produce SBs}
\label{sec:OC_IMF_analysis}

The inflation of a SB is not guaranteed in all open clusters as not all them are able to produce a sufficient number of SNe for that. According to Y17, the   number of SNe necessary for the production of   SBs is $N_{\rm SN, lim} \approx 100$. The mass of the open cluster corresponding to this is 
\begin{eqnarray}
    M_{\rm OC, lim} \equiv N_{\rm SN, lim} / \vartheta \approx 11\,800\msun. \label{eq:MOClim}
\end{eqnarray}%eq:MOClim

\begin{figure}
    \centering
    \includegraphics[width = \columnwidth]{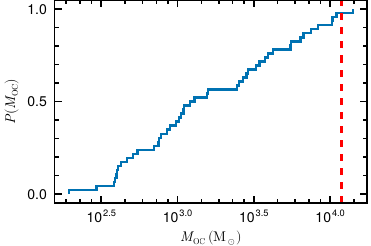}
    \caption{Cumulative distribution of open cluster IMF from \citet{Bhattacharya:2022} used to determine the open clusters able to produce SNe in Sect.~\ref{sec:OC_IMF_analysis}. The vertical dashed line shows $M_{\rm OC, lim}$ defined with Eq.~\eqref{eq:MOClim}}
    \label{fig:OC_IMF}
\end{figure}

We now turn our attention to the fraction of open clusters able to produce a SB and what   their average mass is and the corresponding parameter $s_{\rm r}$ defined in Eq.~(\ref{eq:sr_fiductial}). We take the open cluster IMF ($\equiv \Psi_{\rm OC}$) from the Milky Way (MW) observations. The MW is in a calm non-merger state that produces a basis to study the secular evolution aspects in galaxies. The data for $\Psi_{\rm OC}$ is taken from \citet{Bhattacharya:2022} and is shown in Fig.~\ref{fig:OC_IMF}. 
%fig:OC_IMF

Only one open cluster in the $\Psi_{\rm OC}$ is above the $M_{\rm OC, lim}$ indicating that very few open clusters are able to form a stable SB. The mass of the cluster is $14\,108\msun$, providing the effective value of 
\begin{eqnarray}
    s_{\rm r} = 1.19. \label{eq:sr_value}
\end{eqnarray}%eq:sr_value

In a subsequent analysis, we will need the number of SBs produced per formation of a unit stellar mass ($\equiv \kappa$) that we can derive from the open cluster IMF.  For our  case, the total mass of the \citet{Bhattacharya:2022} catalogue is $137\,514\msun$, indicating that $\kappa = 1 / 137\,514 \approx 7.3\times10^{-6}\,{\rm M_\odot^{-1}}$ SBs are formed per unit stellar mass formation.  %7.271986852247771e-6

%%%

\section{Dynamical coupling of a SB with its environment}\label{sec:analytical_SB}
An evacuated cavity within the interstellar gas introduces a perturbation into the local galactic potential, which alters the trajectories of particles that are making up the ambient medium (e.g. stars and DM). In the case of a non-zero relative velocity between the SB and the ambient medium, this effect resembles the more familiar case of dynamical friction due to a massive object (perturber), by giving rise to an energy--momentum exchange between the perturber and its surroundings.

For the dynamical friction perspective, the main difference between a massive perturber and the SB is that in the latter case, mass is merely displaced from its interior to the SB edges, hence inflating the SB and introducing zero net mass into the environment. 
For the idealised spherically symmetric SB and   Newton's shell theorem, the SB has no dynamical effect on particle trajectories outside its own boundaries. However, within the SB, the same theorem implies that the perturbation to a particle's trajectory arises due to the missing mass within the spherical region interior to the particle's radial coordinate relative to the SB centre, which means that  the particle experiences an acceleration
\begin{align}
\vec{a} = -\frac{GM(<r)}{r^3} \, \vec{r},
\end{align}
where
\begin{align}
M(<r) = \int\limits_0^r 4\pi r'^2 \rho^* {\rm d}r' = - \frac{4\pi r^3 \rho_{\rm env}}{3}.
\end{align}
The last part of the equation holds only for the  interiors of the SBs. 

Compared with the case of a massive object propagating through a medium, there are a few significant differences when considering dynamical friction with a SB as the perturber:
\begin{enumerate}
    \item Since the effective perturbing mass is negative, the particle trajectories tend to diverge rather than converge when traversing the SB. As a result, the wake behind the perturber is underdense rather than overdense relative to the unperturbed state. 
    \item The spatial extent of the SB's dynamical effect is limited to its own size, in contrast to a massive perturber in which case its influence formally extends to infinity. Therefore, no divergence arises when computing the energy--momentum exchange within an extended medium and no factor resembling the Coulomb logarithm appears in the final expression.
    \item In contrast with most massive perturbers, the SB can evolve on timescales of interest for the dynamical friction problem (i.e. the size of the SB at the time of entry and exit of a given particle is generally different). As a result, the encounter is not conservative even in the reference frame of the SB. This effect can dominate the energy exchange compared with interaction with an unevolving perturber, where only momentum is exchanged in the perturber frame (in the limit $m \ll M$). 
\end{enumerate}

The acceleration field for a particle at different times and locations within the evolving SB is shown in Fig.~\ref{fig:accfield_of_bubble}. In keeping with the above discussion, the acceleration peaks near the inner edge of the SB wall where the interior (missing) mass is highest. However, from the perspective of dynamical coupling between the SB and the environment, a more relevant quantity is the net energy gain or loss of a particle during the entire time spent within the SB. We show in Appendix~\ref{app:SB:timedep} that in the frame of the SB, this is determined by the relative radii of the SB at the time of entry and exit of the particle; if $R_{\rm exit} > R_{\rm enter}$, then net energy gain is positive and vice versa (see Eq.~\ref{eq:app:vout}). Since the SB evolves through both expansion and contraction phases, one can see that the dynamical friction force can have either sign, even in a uniform medium, in contrast with a massive perturber.

\begin{figure}
    \centering
    \includegraphics[width = \columnwidth]{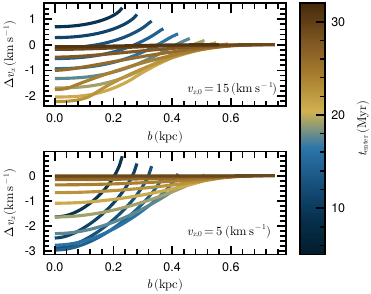}
    \caption{Velocity changes for stars passing through the SB. The top part of the panel shows stars that enter the SB with ${\bf v}_0 = (15,0)\kms$, while in the bottom half ${\bf v}_0 = (5,0)\kms$. The colour depicts the time the star enters the SB (see colour scale at right). }
    \label{fig:dv_through_bubble}
\end{figure}

The change in a star's velocity parallel to its initial direction as it traverses the SB is shown in Fig.~\ref{fig:dv_through_bubble}, for different impact parameters and phases in the SB evolution at the time of entry. At early entry times and for a sufficiently rapidly moving star (upper panel), we have $R_{\rm exit} > R_{\rm enter}$, and the velocity gain is positive regardless of the impact parameter $b$. In contrast, stars that enter the SB in later phases will be able to traverse it only when the latter has already evolved well into the contraction phase. In this case, whether the ratio $R_{\rm exit}/R_{\rm enter}$ is above or below unity depends on the impact parameter, with lower values of $b$ resulting in longer residence times inside the SB, lower values of $R_{\rm exit}/R_{\rm enter}$, and a greater energy loss.

Stars with lower velocities than the characteristic expansion or contraction speed of the SB will in most cases experience a net loss of energy and momentum,\footnote{Here we are referring to stars with velocities directed towards the SB interior, rather than those with outward-directed velocities that are caught up by the expanding SB (which constitute a minority compared with the former in terms of how many enter the SB).} since the SB will have had time to complete the expansion phase and contract to a size $R_{\rm exit} < R_{\rm enter}$ by the time the star exits it (Fig.~\ref{fig:dv_through_bubble}, lower panel).

The significance of the energy gain or loss of a given star compared to its initial kinetic energy is seen from Eq.~(\ref{eq:app:vout}), which can be written as
\begin{align}
\frac{v_{\rm out}^2}{v_{\rm in}^2} = 1 + \frac{ \left[ v_{\rm ch}^2(R_{\rm exit})  -  v_{\rm ch}^2(R_{\rm enter})\right] }{v_{\rm in}^2},
\end{align}    
where $v_{\rm in}$ and $v_{\rm out}$ refer to the stellar velocities at the time of entry and exit, respectively, and $v_{\rm ch}(r) = \sqrt{8\pi G \rho_{\rm env} r^2/3}$. Thus, stars with $v_{\rm in} \gg \max[v_{\rm ch}(R_{\rm enter}), v_{\rm ch}(R_{\rm exit})]$ only obtain a minor correction to their initial energy and vice versa (unless, by chance, $R_{\rm enter} \approx R_{\rm exit}$). The same applies to their trajectories, and hence $v_{\rm ch}$ delineates the boundary between the regime where the phase space density is only slightly deformed by the SB and the collision-like regime, whereby stars are displaced by a considerable distance within phase space.

\subsection{A single SB in an environment}

% \LEt{***One-sentence paragraphs should not be used (even if very long, as this one is). Please check throughout, and
% include single sentences in the previous or following paragraph, as appropriate, or
% rephrase (i.e. add at least one more sentence). }
The total momentum transfer between a SB and a stellar population is found by double integration. First, over the phase space of stars and second, the SB lifetime. The outcome can be expressed as
\begin{align}
\Delta\vec{P} = \int \Delta\vec{p} \, {\rm d}F_{\rm in} \, {\rm d}S_{\rm b} \, {\rm d}t_{\rm in},
\label{eq:dP}
\end{align}
where $\Delta\vec{p}$ is the momentum transfer from a single encounter, $dF_{\rm in}$ is the flux of stars through the unit area of the SB surface, $dS_{\rm b}$ is the surface element (see Eqs.~(\ref{eq:app:dF2}) and (\ref{eq:app:dS})), and $t_{\rm in}$ is the entry time.

We note  that in general the stellar flux $F_{\rm in}$ through the SB surface depends on the location on this surface; obtaining a non-zero value for $\Delta\vec{P}$ relies on this being the case (otherwise the integral in Eq.~(\ref{eq:dP}) vanishes by symmetry). Typically, this reflects the relative motion between the SB and the stellar component, whereby more stars are encountered on one hemisphere of the SB.
If there exists a frame in which the stellar distribution is approximately homogeneous and isotropic, then $\Delta\vec{P}$ is non-zero only if the SB is moving relative to this frame. From symmetry considerations one finds that the momentum transfer $\Delta\vec{P}$ is parallel to the direction of the SB's motion in this frame. In contrast with the classical massive perturber, however, $\Delta \vec{P}$ along this direction can take either sign depending on the specific parameters.

To illustrate the momentum exchange between the SB and the stellar population, we consider an idealised configuration in which a spherical SB with a well-defined (thin) edge moves through a homogeneous stellar field with a Maxwellian velocity distribution. This  set-up allows a semi-analytic treatment outlined in Appendix~\ref{app:SB:PhSp}, which can serve as a validation of the full numerical model. 

\begin{figure}
    \centering
    \includegraphics[width = \columnwidth]{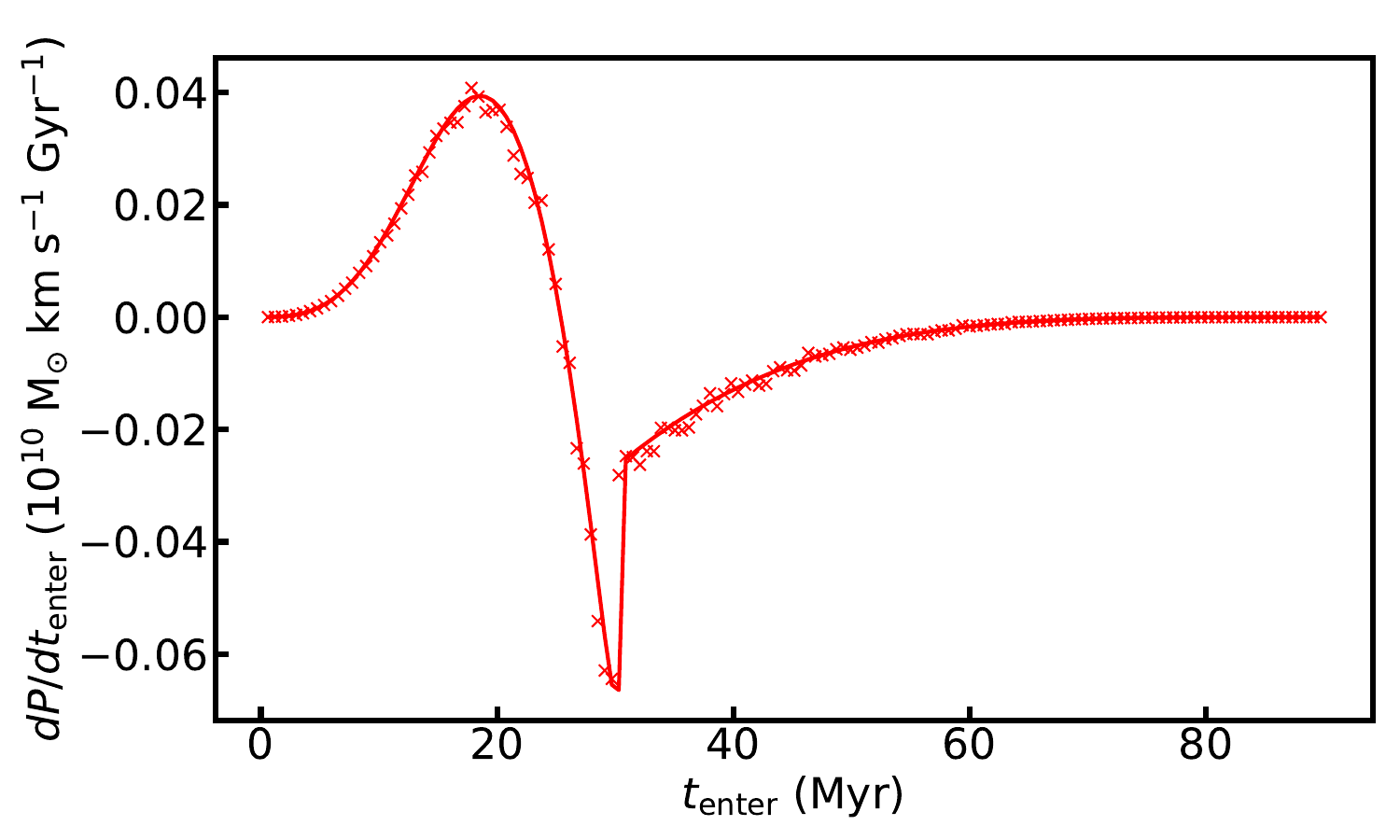}
    \caption{Momentum exchange rate between the stellar population and an idealised thin-edged SB with constant expansion and contraction velocities $v_{\rm exp} = 20$~km~s$^{-1}$ and $v_{\rm col} = 10$~km~s$^{-1}$, respectively, and expansion phase duration of 30~Myr. The ambient gas density is $0.015\,{\rm M_\odot\,pc^{-3}}$. The stellar density is $0.1 \,{\rm M_\odot\,pc^{-3}}$, with a Maxwellian velocity distribution with $\sigma = 25$~km~s$^{-1}$. The relative velocity between the SB and the stellar distribution is $v_{\rm b} = 10$~km~s$^{-1}$. The solid line shows the semi-analytical solution given by Eq.~(\ref{eq:app:dPdt2}); the symbols correspond to the full numerical solution.}
    \label{fig:dPdt}
\end{figure}

Figure~\ref{fig:dPdt} shows the momentum gain experienced by stars as a function of their entry time, multiplied by the rate at which they are entering the SB at that time; the total gain or loss $\Delta\vec{P}$ over the SB lifetime is found by integration over $t_{\rm enter}$. In broad terms, the stellar population experiences a net momentum gain (along the direction of the SB propagation) in the early phases of the SB evolution and vice versa. This is consistent with the discussion in the previous section, whereby a star entering the SB in its growing phase will more frequently have $R_{\rm exit}/R_{\rm enter} > 1$ and gain energy by the encounter. As expected, most of the momentum exchange is done by stars that traverse the SB when the latter is not far from its  maximum size, while stars entering either very early or very late in the SB life cycle have $R_{\rm enter}, R_{\rm exit} \ll R_{\rm max}$ and their energy--momentum transfer is correspondingly small (see Eqs.~\ref{eq:app:vff} and \ref{eq:app:vout}).

Finally, the angular momentum can be acquired in the same way as momentum. The integration is done over the angular momenta changes of individual particles ($\Delta L_z'$):
\begin{align}
\Delta L_z = \int \Delta L_z' \, {\rm d}F_{\rm in} \, {\rm d}S_{\rm b} \, {\rm d}t_{\rm in}.
\label{eq:dLz}
\end{align}

\subsection{Including all the SBs}

Superbubbles do not usually occur as single isolated entities; instead, in most star-forming galaxies, they are generated at a certain rate throughout the gas disc. The resulting dynamical coupling between the stellar and gas components can be described by some average momentum exchange rate between the two components per unit disc area. Specifically for the case of circular orbits within the gas disc, the exchange can be characterised in terms of either linear or angular momentum, collectively denoted as $\Delta\Theta = \{\Delta P_x, \Delta L_z\}$.

\subsubsection{Transfer of linear momentum}\label{sec:lin_momentum_transfer}

First, we consider the exchange of linear momentum, taking $\Delta\Theta = \Delta P_x$. Within a disc-like environment, we consider a gas slab with area $A$ and height $h$, and denote the rate of SB formation per unit area as $f_{\rm b}$. As the star formation activity is concentrated towards the mid-plane of the gas disc (the region of coldest gas), we assume that the SB centres are likewise in the mid-plane, disregarding any vertical variation.

The rate of SB formation in the above region is therefore $f_{\rm b} A$, each SB contributing an exchange of momentum $\Delta P_x$. This yields a momentum gain rate of gas within the region $-f_{\rm b} A \Delta P_x$ and a corresponding rate for the stellar population (with opposite sign), or equivalently, an effective force per unit area, $dF_x/dA = -f_{\rm b} \Delta P_x$.

The gas mass per unit area (i.e. surface density) is $\Sigma_{\rm gas} = \rho_{\rm gas} h$. Hence the gas deceleration rate is
\begin{eqnarray}
s \equiv \frac{\mathrm{d}v_{\rm gas}}{\mathrm{d}t} =
-\frac{1}{\rho_{\rm gas} h} \frac{{\rm d}F_{\rm x}}{{\rm d}A}
=-\frac{f_{\rm b} \Delta\!P_x}{\rho_\mathrm{gas} h }\label{eq:s_definition}.
\end{eqnarray}

\subsubsection{Transfer of angular momentum}
The angular momentum gain or loss of the gas disc can be evaluated by taking $\Delta\Theta = \Delta L_z$; the effect can be characterised by the relative rate of change of specific angular momentum of a gas element, or equivalently, the rate of radial migration of a disc annulus.

The former essentially yields a timescale (more precisely, its inverse) in which a  significant change in  angular momentum  can take place,%\LEt{***have I\ interpreted correctly? }

\begin{eqnarray}
\zeta \equiv \frac{\dot{L}_z}{L_z} = -\frac{\Delta L_z f_{\rm b}}{\Sigma_{\rm gas} R v_{\rm c}},\label{eq:zeta_def}
\end{eqnarray}%eq:zeta_def
where $R$ is the galactocentric distance, and we have assumed that the rotational velocity is very close to the circular speed $v^2_{\rm c} = -R\partial \Phi/\partial R$ determined by the galactic potential.

To determine the resulting radial migration speed, one can use the continuity and angular momentum conservation equations for the gas disc in cylindrical coordinates,
\begin{align}
    \frac{\pd}{\pd t} \Sgas + \frac{1}{R}\frac{\pd}{\pd R} \left( R \Sgas \vr \right) = 0
    \label{eq:cont}
\end{align}
and
\begin{align}
    \frac{\pd}{\pd t} \left(\Sgas R^2 \Omega\right) + \frac{1}{R}\frac{\pd}{\pd R} \left( R\Sgas \vr R^2 \Omega \right) =
    %\frac{1}{2\pi R} \frac{\pd G}{\pd R},
    T_{\rm A},
    \label{eq:angmom}
\end{align}
respectively,
where $\vr$ is the radial velocity and the right-hand side is the imposed torque per unit surface area. The parameter $\Omega$ denotes the angular velocity. Using Eq. (\ref{eq:cont}), the angular momentum equation can be rewritten in the form
\begin{align}
    \vr \frac{\pd}{\pd R} \left( R^2\Omega \right) =
    %\frac{1}{2\pi R \Sgas} \frac{\pd G}{\pd R}
    \frac{T_{\rm S}}{\Sgas}
    \equiv T,
    \label{eq:angmom2}
\end{align}
where $T$ is the torque per unit mass, and we   assume $\pd (R^2\Omega)/\pd t = 0$, which is justified if the orbital velocity only depends on $R$.

% \LEt{***paragraph }
In the present context, the torque per unit mass from dynamical friction is $T = \Delta L_z f_{\rm b}/\Sigma_{\rm gas}$, whereby Eq.~(\ref{eq:angmom2}) yields 
\begin{align}
\vr = \frac{\Delta L_z f_{\rm b}}{\Sigma_{\rm gas} \,\partial (R^2\Omega)/\partial R}
= \zeta R \left[ 1 + \frac{{\rm d}\ln v_{\rm c}}{{\rm d} \ln R} \right]^{-1},
\end{align}
where we have used  $v_{\rm c} = R\Omega$.

Although the inward speed is a very intuitive quantity, a more frequently used measure of gas migration is the flux of mass through some radius. The mass transfer inwards is included by converting $v_R$ to mass flux by multiplying it with surface density and the area shifted over the time interval:
\begin{equation}
    {\rm d}M_{\rm gas} = \Sigma_{\rm gas}2\pi R v_R{\rm dt}.\label{eq:mass_transfer}
\end{equation}

\subsubsection{The SB formation rate}\label{sec:bubblerate}
Both SB influence estimators, \eqref{eq:s_definition} and \eqref{eq:zeta_def}, depend on the SB rate, to which we provide three estimates: from direct observations, from the Kennicutt--Schmidt relation and from the SB saturation limit. 

\citet{Ehlerova:2013} measured the surface density of SBs with radius over $100\pc$ in the solar neighbourhood to be about  $4$ per ${\rm kpc^2}$. To convert this into the  SB rate, we use the lifetime of SB $t_{\rm end} = 71.6 \Myr$ (see Sect.~\ref{sec:bubble_evolution}), from which during $31.6\Myr$ it has an inner radius exceeding $100\pc$. To have four superbubbles per square kiloparsec, the formation rate of SBs should be $f_{\rm b} = 127\unitbubblerate$. 

The Kennicutt--Schmidt relation can be used for SB rate estimates as well. In Sect.~\ref{sec:OC_IMF_analysis}, we defined the conversion rate ($\kappa$), showing how many SBs form per formed stellar mass. Hence, we can convert the surface density of the star formation rate to the SBs formation rate by simple multiplication. Using the Kennicutt--Schmidt relation for non-starburst galaxies from \citet{Reyes:2019} with the surface density of gas  $13.6\,{\rm M_\odot pc^{-2}}$\citep{McKee:2015} we obtain the rate of SBs $f_{\rm b} = {42\unitbubblerate}$. 

\begin{figure}
    \centering
    \includegraphics{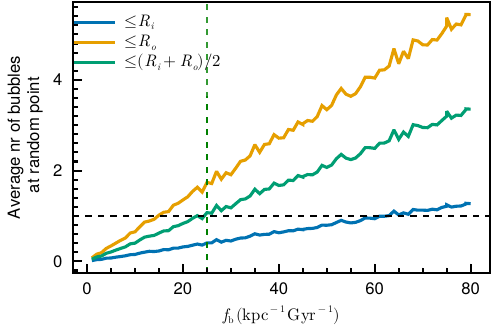}
    \caption{Average number of SBs that are in a random point based on simple numerical simulation. }
    \label{fig:bub_rate_saturation}
\end{figure}

If the formation rate of SBs is high and the lifetime of SBs long, they start to overlap, and new ones   form  on the edges of older SBs \citep{Barnes:2022}. This does not allow the SB to complete its full evolution cycle, and our approximation of a single SB influence breaks down. To estimate the SB rate when the saturation of independent SBs occurs, we assume that a higher SB rate does not produce a larger $\Theta$ transfer. We derive the saturation limit by simulating the formation of SBs at random positions and times, and $f_{\rm b}$ measures the number of random points inside of SBs (defined as the distance from a random point to the SB centre is less than $R_{\rm m}$) on average. At a SB formation rate of $f_{\rm b} = 25\unitbubblerate$ the number of overlapping SBs will be over unity on average, and we use this rate as the saturation limit. Figure~\ref{fig:bub_rate_saturation} shows the saturation level of our simulation. 
% \end{itemize}

Considering that all these estimates are diverse and almost independent, we choose the most conservative one for further analysis: $f_{\rm b} = 25\unitbubblerate$. Changing the value of $f_{\rm b}$ changes the secular evolution speed proportionally. 

\section{Examples of SBs in various environments}\label{sec:exmamples}
\subsection{SBs in a homogeneous isotropic environment}\label{sec:bubble_like_Chandrasekhar}
This section shows the amount of de-acceleration ($s$) SBs induce in gas. We assume a homogeneous and isotropic environment as it is an approximation to a more realistic environment.
First, we solve the SB influences numerically using \eqref{eq:s_definition}, and then construct an approximation formula to simplify its future evaluations. 

The set-up of the environment is a slab of gas, which we complement with homogeneous stellar distribution with isotropic velocities. The parameter $v$ denotes the average velocity between the gas and stars, an analogue to asymmetric drift; $\sigma$ denotes the stellar velocity dispersion. We described the calculation prescription  in  Sect.~\ref{sec:lin_momentum_transfer}, and the implementation parameters are in Table~\ref{tab:Ch_analog}.
\begin{table}
    \centering
    \caption{Parameters for determining the level of the influence to the gas slab by the stars via SBs. }
    \label{tab:Ch_analog}
    \begin{tabular}{l|l|l}
    \hline
         & Default value & Comment\\
        \hline
        $v$ & $10\kms$ & \makecell[l]{The velocity of SB with respect to gas \\This and the next parameter are used in \\phase space density in Eq.~\eqref{eq:f_normalisation}} \\
        $\sigma$ & $20\kms$ & The velocity dispersion of the stars \\
        $v_{\rm col}$ & $10\kms$ & Sets the SB collapse in Eqs.~\eqref{eq:Ri_fid} and \eqref{eq:Ro_fid}\\
        $s_{\rm r}$ & $1.19$ & \makecell[l]{SB scaling defined in Eq.~\eqref{eq:sr_def}. With \\ $v_{\rm col}$, it fixes SB lifetime $t_{\rm b} = 71.58\Myr$}\\
        $\rho_\star$ & $0.1\,{\rm M_\odot\,pc^{-3}}$ & Stellar density used in \eqref{eq:f_normalisation}\\
        $\rho_{\rm gas}$ & $0.015\,{\rm M_\odot\,pc^{-3}}$ & Density of the gas hosting SB \\
        % $\rho_{\rm gas}$ & $0.0535433\,{\rm M_\odot\,pc^{-3}}$ & Density of the gas hosting superbubble. \\
        $f_{\rm b}$ & $25\unitbubblerate$ & The rate SBs form; see Sect.~\ref{sec:bubblerate}\\
        \hline
    \end{tabular}
\end{table}
\begin{figure}
    \centering
    \includegraphics[width = \columnwidth]{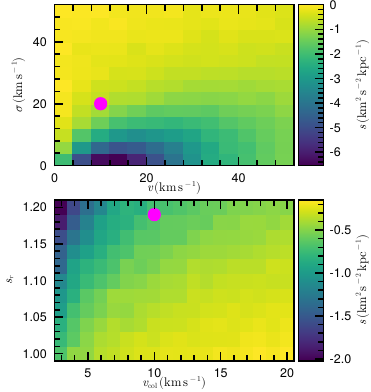}
    \caption{De-acceleration $s$ in a homogeneous environment. The long-term effects can be estimated using the relation $\kms$ per ${\rm Gyr}$ as it approximates $\accunit$. The top panel shows the dependence of $s$ on the kinematical parameters of the homogeneous environment: the velocity and velocity dispersion. The magenta point shows the difference between MW gas and stellar disc values. The bottom panel shows the dependence on the SB parameters of collapse speed and scaling, with kinematic parameters $v$ and $\sigma$ fixed at the magenta point in the top panel. The de-acceleration field can be approximated using Eq.~\eqref{eq:ch_approx}. 
    }
    \label{fig:slowdown_2panels}
\end{figure}
% fig:slowdown_2panels

Figure \ref{fig:slowdown_2panels} shows the resulting de-acceleration of the gas slab, both in kinematic and SB property aspects. 
For the context, MW-like galaxy that formed its discs around redshift $1.5-2$ \citep{vanderWel:2014} giving them the time-span to modify their discs for about $10\Gyr$ once they evolve to redshift $0$. 
% \LEt{***at about 10 Gyr, by 10 Gyr, after 10 Gyr? Also the following sentence seems to be missing something (migrate from what?). Please rephrase. } 
If we assume a flat rotation curve with a velocity plateau of $230\kms$, then during that time the gas can migrate from due to SB $\approx10s/230$, or $20\%$ if $s=5\accunit$. The evaluation assumes a homogeneous host environment, which we discard in the next section. 
Figure \ref{fig:slowdown_2panels} shows that the highest impact comes from the parameter space where the stellar disc is cold, its velocity dispersion is below $10\kms$, but with some dependence on the relative velocity of the gas disc and the stellar disc. The bottom panel show the dependence on the SB properties, with slow-collapsing SBs being more efficient in transferring momentum:  a longer age extends the time of momentum transfer. Analogously, the larger size (described by  $s_{\rm r}$) increases the cross-section of the SB, allowing more stars to transfer their momenta. Larger SBs also have larger amounts of  displaced gas mass and they provide higher accelerations (see Sect.~\ref{app:bubble_calc_formulae}).

To evaluate the de-acceleration with more ease, we fitted an empirical approximation formula to the top panel in Fig.~\ref{fig:slowdown_2panels}. We made an analytic approximation for the simple case using the SymbolicRegression package \citep{Symbolicregression} in Julia \citep{Julia-2017}. The mean and standard deviation of the residuals of the fit is  $-0.0013\accunit$,  $0.257\accunit$. The best model has the form
\begin{equation}
    s_{\rm approx} = \frac{v_{\rm bub} + \sigma + a_1}{ a_3(v_{\rm bub} - a_2) \left[ a_5 + \exp\left(\frac{\sigma + a_4}{v_{\rm bub}}\right) \right] },\label{eq:ch_approx}
\end{equation}
where $a_1 = -112 \kms$,
$a_2 = -1.64\kms$, 
$a_3 = 0.650\accunit$,
$a_4 = 5.94\kms$, and
$a_5 = -0.217$. This equation assumes that the velocities are in units of $\kms$, distances in $\kpc,$ and the resulting de-acceleration in $\accunit$.

\subsection{SBs in a test galaxy}
A homogeneous environment can approximate regions without a steep density gradient and host acceleration field. However,  the orbits in a galaxy can have a circular nature, and they can produce mismatches with the approximation. In this section we characterise a situation where $\Theta = L_z$, describing the SB effects by changes in angular momentum. 

\subsubsection{Underlying galaxy}\label{app:host_description}
The numerical estimations of the order of magnitude effects require specifying a galaxy. We use a simple analytical potential for a dynamical model evaluation that mimics a disc galaxy.
% \LEt{***as written it says that you, the authors, resemble a galaxy. Please rephrase. \    } 

The galaxy has two density components: DM is described with a Navarro-Frenk-White (NFW) profile \citep{NFW:1996}, and stellar distribution is described with Miyamoto-Nagai (MN) profile \citep{MN:1975}. The gas disc is assumed to produce an insignificant contribution to total density, and hence, its contribution to gravitational potential is neglected. The MN profile has the potential 
\begin{eqnarray}
    \Phi_{\rm MN} = \frac{-GM_{\rm MN}}{\sqrt{R^2 + \left[a_{\rm MN}+\sqrt{z^2 + b_{\rm MN}^2}\right]^2}} \label{eq:MN}
.\end{eqnarray}
The chosen values for the  MN profile were $a_{\rm MN} = 3.0\kpc$, $b_{\rm MN} = 0.4\kpc$, and $M_{\rm MN} = 7.11\times10^{10}\msun$. The NFW profile has density in the form of
\begin{eqnarray}
    \rho = \frac{\rho_{\rm NFW}}{\frac{r}{R_{\rm NFW}} \left( 1 + \frac{r}{R_{\rm NFW}} \right)^2} \label{eq:NFW}
\end{eqnarray}
with the parameter values $\rho_{\rm NFW} = 0.001\times{\rm 10^{10}M_\odot\,kpc^{-3}}$ and $R_{\rm NFW} = 15\kpc$. 

To evaluate the SB influence on the galaxy, we need to know the stellar and DM particle kinematics. They are found from the Jeans equations \citep{Jeans:1915} by assuming stationarity and cylindrical or spherical symmetry for the disc and DM halo, respectively. 

The first equation for solving disc kinematics is the radial Jeans equation
\begin{eqnarray}
    \frac{\partial (\rho\sigma^2_R)}{\partial R} = -(1-\beta^2)\rho\frac{\partial\Phi}{\partial R} - \frac{1-\alpha^2}{R}(\rho\sigma^2_R),
\end{eqnarray}
with the simplifying assumptions that tangential dispersion is proportional to the  radial dispersion and rotational velocity to circular velocity:
\begin{eqnarray}
    \sigma_\theta &=& \alpha \sigma_R\label{eq:def_alpha},\\
    v_{\rm rot} &=& \beta v_{\rm circ}\label{eq:def_beta}.
\end{eqnarray}
In our applications the velocity ellipsoid is slightly elongated in the radial direction, $\alpha = 0.8$, and the asymmetric drift (the relative speed between the circular  and rotational velocities) is about $\approx10\kms$ as we chose $\beta = 0.95$. The vertical motions originate from solving the vertical Jeans equations by assuming that the tilt term in Jeans equations is zero:
\begin{eqnarray}
    \frac{\partial (\rho\sigma^2_z)}{\partial z} = -\rho\frac{\partial \Phi}{\partial z}.
\end{eqnarray}
We solved the equations using the Runge-Kutta-Fehlberg method by integrating $\rho\sigma_r^2$ or $\rho\sigma^2_z$ along the coordinate axis\footnote{The characteristic aligns with coordinates.} and then solving for the kinematical property. 

\begin{figure}
    \centering
    \includegraphics[width = \columnwidth]{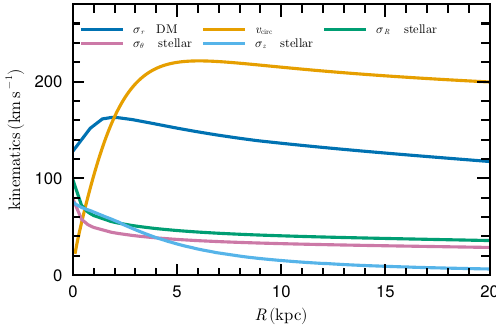}
    \caption{Kinematic solution of the test-galaxy from solving the Jeans equations. The rotation curve of the disc is $\beta = 0.95$ times the circular velocity curve. }
    \label{fig:vcirc_curve}
\end{figure}

By assuming non-rotating isotropic velocity distribution, the solution for the spherical Jeans equations for the DM kinematics was provided in \citet{Kipper:stream_asymm}. Our aim is also to test if the rotation of DM influences the angular momentum transfer by gas. We wanted to achieve a situation where as many other parameters as possible (aside from the rotation) are shared with the non-rotating solution. The rotation can be induced by modifying the spherical solution by lowering one side of the $v_{\rm rot}$ distribution ($p_{\rm rot,\,DM}$) on one side and increasing it by the same amount on the other side, similar to what was done in \citet{joshi2023formation}. Effectively, it means that the overall potential and density remain the same if the rotational direction of a fraction of stars is switched.\footnote{Technically, this produces a valid solution for the solved point, but loses the validity of isotropy in other radii. For the current application it is a suitable approximation.} The factor $p_{\rm rot, DM}$ describes the modification of the distribution function
\begin{eqnarray}
    f_{\rm rot} = \frac{1}{1 + \exp(-kx)} - \frac12,
\end{eqnarray}
with $x\equiv |v_{\rm rot}|/\sigma_{\rm DM}$. The value of  $p_{\rm rot, DM}$ is multiplied by $1 \pm f_{\rm rot}$, depending on whether $v_{\rm rot}$ is positive or not. In the implementation,  $k$ equals $\pm5$, which transforms the mean value to $\pm0.37\sigma$. The solution of the kinematics is in Fig.~\ref{fig:vcirc_curve}.

\subsubsection{The gas disc changes}\label{sec:changes_in_disc}
\begin{figure}
    \centering
    \includegraphics[width = \columnwidth]{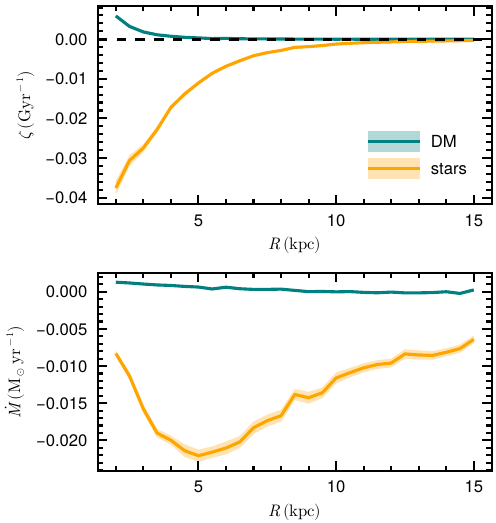}
    % \caption{The dependence of gas disc expansion due to SBs both in terms of relative angular momentum change (top panel) that is defined with Eq.~\ref{eq:zeta_def} and expansion speed (bottom panel). 
    % }
    % \includegraphics[width = \columnwidth]{Figs_revision1/R_dep_with_gas_transport.pdf}
    \caption{Dependence of gas disc expansion due to SBs   in terms of relative angular momentum change (see Eq.~\eqref{eq:zeta_def}) and of mass transfer (see Eq.~\eqref{eq:mass_transfer}) in the top and bottom panel, respectively.  %\ET{It is unclear what the bold text is saying? If I understand correctly then bottom panel shows the gas mass transfer computed using Eq. 18.} 
    }
    \label{fig:gas_disc_expansions}
\end{figure}

Figure \ref{fig:gas_disc_expansions} shows the influence of the SBs on the gas disc in two representations: relative loss of angular momentum and radial velocity.  
We note two important things here. Firstly, there is a strong radial dependence in the inner region of the galaxy. The innermost region was not calculated as the SB would not fit the region without breaking down the shape assumptions. Secondly,  the contributions by the DM and stellar components have opposite signs. 
% \LEt{***have I\ interpreted correctly? (I needed to make a complete sentence). } 
This is specific only to the region where particles pass through the SB very quickly and the gravitational potential is cylindrical and does not occur in the homogeneous and isotropic medium. There are multiple effects in play simultaneously, so pinpointing the exact culprit for this effect is not feasible in the current study. 

\begin{figure}
    \centering
    \includegraphics[width = \columnwidth]{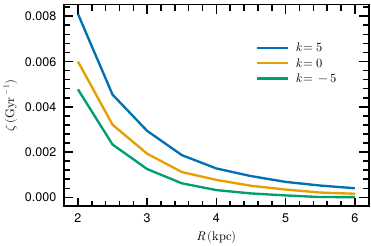}
    \caption{DM rotation influence to the transfer of angular momentum between DM and the gas disc ($\zeta$). Different $k$ values correspond to different rotations, with value $\langle v\rangle = 0.37\sigma$, the dispersion of the DM halo is shown in Fig.~\ref{fig:vcirc_curve}.}
    \label{fig:DM_rot}
\end{figure}

The rotation of the DM profile might produce some changes in the stellar disc (e.g. \citealt{Herpich:2015, Kataria_2022, Li_2023}). One physical reason is that the dynamical friction is sensitive to the kinematics of the particles, not only density \citep{Chandrasekhar:1943}. Therefore, we investigated the role of DM rotation on the gas disc via the dynamical friction. Figure~\ref{fig:DM_rot} shows the difference between different DM rotation models. We conclude that the contribution from DM kinematics exists, but that it is weak. 

\section{Discussion}\label{sec:discussion}
\subsection{Observational verification of the SB profile}\label{sec:YadavLineOfSight}
In Sect. \ref{sec:density_profile} we describe the density profile of a SB as derived in Y17. 
In this section we   verify the density's consistency with the observations.   \citet{Watkins:2023} analysed the galaxy NGC 628 in terms of its SB content from the observations from the PHANGS46–JWST using   MIRI’s F770W filter. We took advantage of this observation and focused on the SB situated at approximately $3~{\rm kpc}$, south-east from the centre of this galaxy. This SB is an ideal example of these processes as it has an almost circular shape.

We measured the surface brightness in circular regions from the centre of the SB, matching our coordinate centre. The Y17 results assume initially homogeneous gas distribution, but in our case, this is not fully satisfied if we want to investigate the ISM regions further than the extent of the SB. To test if this was an issue, we measured the surface brightness only in a well-defined shell-like structure as seen in the JWST image, also done in the work of \citet{Mayya:2023}. This approach accounts for the influence of various external factors, including the collapse of edges, gas diffusion, and other effects.

We generated the intensity of the profile from the intensity map of the image by averaging intensities at different radii from the shell centre. Thereafter, we binned these radii so that we could get a smoother profile.  For the binned data, we calculated the mean and the error of the mean by bootstrapping \citet{bootstrap_mainpaper}.  This gave  us the observed surface brightness profile within errors, which is proportional to the density $(\rho)$ profile given by Y17.
 The theoretical density profile is approximated in Sect. \ref{sec:density_profile} following the Y17 simulation. This simulation assumes a homogeneous environment for all the initial conditions. The observations are more complex as the ISM and gas outside the SB are not homogeneous as well, and there are other effects contributing, such as the higher gas density on the spiral where one side of the SB resides or some higher rates of star formation at the edges of the SB from the pressured gas or other specific gas properties. We denote this conversion factor between the surface density of the Y17 fiducial profile and the surface brightness of the observations by $C$.  To exclude ISM effects in the observations, we take into account radii up to $0.7$ kpc, a bit further than the well-defined luminosity shell of the profile. Additionally, the observations have a background intensity that we denote as $B$.

The conversion between the line of sight and SB-centric coordinate in a galaxy is $l = \pm\sqrt{r^2 - R^{2}}$, where $r$ is the 3D distance from the centre of the SB and $R$ is the distance from the centre of the circle as measured in the photometric image.
Thus, the line-of-sight integral is defined as
\begin{eqnarray}
  I(R) &=& C \int_{-l_{\rm lim}}^{+l_{\rm lim}} \rho[r(l)] \, {\rm d}l + B 
.\end{eqnarray}
This gives us five free parameters, three that are the result of the observational image, $C$, $B$, $l$, and two that describe the SB, its expansion time $t$ and the scaling factor $s_r$ (analytically described in Sect.~\ref{sec:bubble_evolution}). These parameters are modelled within large and uniform priors. Our aim was to match the observed data with the theoretical model by following a Bayesian approach \citep{Laplace1774}.
We considered errors as coming from the standard deviation derived from bootstrapping plus a systematic uncertainty to account for the fact that the SB is not isolated or in a homogeneous environment.

The following equation gives us the likelihood of the posterior. We used $10^5$ iterations in an adaptive Monte Carlo scheme provided by Julia, as implemented by \citet{MCMCVihola:2020}: 
\begin{eqnarray}
\log \mathcal{L}  \propto  -0.5  \sum_{i=1}^{N} \left[ \left( \frac{{I_{\text{obs},i} - I_{\text{model},i}}}{{\Delta I_i}} \right)^2 \right] \label{eq:likelihood_cos}
.\end{eqnarray}
Here $I_{\text{obs}}$ denotes
the surface brightness  coming from the JWST image in the radii bins and
the Markov chain\ Monte Carlo inference estimates of the free parameters
for the $I_{\text{model}}$, and $\Delta I_i$ represent the associated errors.

 The constraints of the time of the expansion of the SB to $t_{\rm obs} = 0.030 \pm 0.002$ Gyr indicate that the time of the observation of the specific SB's evolutionary stage is at its biggest peak before starting to shrink. We also constrained the scaling factor to $s_r = 1.18 \pm 0.04$, translating to $ \approx 230$ progenitor SN explosions. For the rest of the parameters, we obtained the line-of-sight distance $l_{\rm lim} = 0.37 \pm 0.02\, { \rm kpc}$ and the luminosity background information of the JWST image as $B = 10.79 \pm 0.03\, {\rm Jy/sr}$. The latter is within the measurements consistent with regions outside the galaxy. 

\begin{figure}
    \centering
\includegraphics[width = \columnwidth]{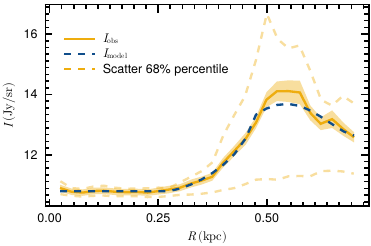}
    \caption{Comparison between the surface brightness from the  JWST image and the surface brightness   from the Y17 best fit of the model.  The observational flux of the SB (gold line) is plotted between three sigma (light gold dashed lines) and the standard deviation plus the systematic error of the bootstrap analysis for the binned data. The best fit of the model incorporating the five free parameters is superimposed on them (blue dashed line)}
    \label{fig:lineoofsight}
\end{figure}

As demonstrated in Fig.~\ref{fig:lineoofsight}, the theoretical model closely matches the observational data, providing an agreement between the approach of the Y17 profile and the observed data. This verifies that the theoretical density profile from Y17 describes accurately the SB phenomena.

\subsection{Robustness of the assumptions and approximations} \label{sec:robustness_of_assumptions}
Our analysis relies on a few assumptions about the SB itself, its environment and the stars interacting with it. The SB description is based on the Y17 simulation and their assumptions about the SB propagate here. Their dominant assumptions are homogeneous initial gas distribution with gas density analogous to the solar neighbourhood and the stars exploding near the centre being distributed uniformly in time. 

\subsubsection{Environment assumptions}
The gas distribution in real galaxies is not homogeneous; the radial and vertical gradients of the gas density contradict our assumptions. The S\'ersic profile describes the HI surface brightness distribution, whereas the S\'ersic index characterises the steepness or gradient of the profile. Low values indicate nearly constant core with truncation, and large values ($\sim 4$) indicate cuspy profiles. The S\'ersic index of averaged matter distribution of spirals from the THINGS survey is between $0.18 - 0.36$ \citep{Brinks:2017}, showing the gradient of the gas disc to be much shallower than a typical stellar disc (S\'ersic index $\sim 1$), and thus the homogeneity assumption in the radial direction is a good approximation. The vertical gradient of gas discs is much steeper. The scale heights of different types of gas vary between $75\pc$ for molecular gas to $3000\pc$ for a hot ionised medium \citep{Ryden:2021}. The highest contribution comes from a warm neutral medium with a scale height of $\sim300\pc$, which corresponds to the fiducial SB inner radius at the maximum extent. To see if the disc-height limited profile still matches the theoretical profile, we modelled observations with our  SB model. In  Sect.~\ref{sec:YadavLineOfSight} we show that the profiles match best when the integration extent over the line of sight is $370\pc$, which slightly exceeds the scale height of the gas disc, indicating that the SB almost fits the gas discs. 

We intrinsically assume that the shape of the SB remains spherical. One reason for deviations from sphericity is shearing motions from galactic rotation. We tested the extent of shear distortions by assuming a flat rotation curve with $v_0$, the SB is at galactocentric radius $R_0$, the radius of SB is $R_{\rm SB}$, and the time of shearing is $\Delta t$. In that case, the far edge of the SB can shift by $v_0R_{\rm SB}\Delta t/R_0\approx 225\pc$ from the centre when assuming values $\Delta t=30\Myr$, $R_0=8\kpc$, $R_{\rm SB}=300\pc$, and $v_0 = 200\kms$.\footnote{ The calculations assume there are no compensating forces (e.g. pressure gradients) to restore the round shape.} When trying to see whether the elongation and tilt are present in observations (e.g. the \citealt{Barnes:2023} images), we can see it for some SBs, which is a plausible deviation from our assumption. However,  \citet{Binney:2008} showed that the gravitational potential ellipticity is of the order of  one-third to one-half of the density ellipticity. Our current example has an ellipticity of $0.43$, which translates to a potential ellipticity of $0.14- 0.23$ or an axis ratio of $\approx 0.82$. Considering other uncertainties of the SB rate, evolution, and structure precision, we do not consider this deviation to be of first-order importance.

The secular evolution of the gas disc due to SBs is proportional to the rate of SB formation in the galaxy (see Eqs.~\eqref{eq:s_definition} and \eqref{eq:zeta_def}). Although our estimates of the SB formation rate have a high scatter (see Sect.~\ref{sec:bubblerate}), we would like to point out the possibility that there are spatial variations of the rate within a galaxy and between the galaxies. \citet{finn2024almalegus} show that most massive molecular clouds, producing a greater number of  likely open clusters able to develop into SBs, occur more likely in prominent spirals, contrary to inter-arm regions or regions in flocculent galaxies. In the latter cases, we suspect that our de-accelerations are overestimated. However, although smaller SBs   also have smaller effects on the galaxy evolution (see Fig.~\ref{fig:slowdown_2panels}), there are more of them and the cumulative effect can be comparable.

\subsubsection{SB assumptions}\label{sec:discussion_SB_assumptions}
In the case of the SB physics, we assume that the shape of the SB is the same as Y17 described, and that it has the same scaling relation they provided. We describe  the suitability test of the profile in Sect.~\ref{sec:YadavLineOfSight}. 

Although the observed and theoretical profiles match, the link between the SB and the SNe/SFR needed to create them needs a discussion. In Sect.~\ref{sec:OC_IMF_analysis}, we adopted the approach that $100$ SNe produce a stable SB and evaluated the results accordingly. Fewer SNe are also likely to make a SB, but it would not be very stable. The smaller number of SNe can be significant; for example,  \citet{Barnes:2022} showed many SBs in a galaxy when observed at high resolution. We suspect the smaller ones can also transfer angular momentum, but less effectively (see Fig.~\ref{fig:slowdown_2panels}). This paper does not provide the analysis for that. 

Figure \ref{fig:slowdown_2panels} shows that in addition to the size of the SB, the collapsing speed ($v_{\rm col}$) strongly influences how much the SB affects the transfer of momentum. If the corresponding time of the SB is twice as fast, then the transfer efficiency reduces, but the SB rate ($f_{\rm b}$) increases.\footnote{Here we are assuming that the limiting factor of SBs is the spatial saturation (see Sect.~\ref{sec:bubblerate}).} 
\citet{Kinugawa:2023} demonstrated that due to mass transfer between binaries, the time when the SNe can occur in an open cluster gets extended, possibly delaying the collapse of the SBs. 

Superbubbles are destroyed in simulations due to shearing motions within the disc and/or are mixed into the ISM by hydrodynamical instabilities and turbulence, for example. In our analytical approach we cannot mimic it well, but we included an analysis to see what would happen to the transfer rate when the edges diffuse instead of collapse (i.e. when the inner and outer edge of the profile drift apart until the inner radius reaches zero). This analysis is provided in Appendix~\ref{app:SB_diffusion}. 

\subsection{Implications}
\citet{DellaCroce:2023} reported that the outskirts of open clusters are expanding, the outer parts faster than the inner. The trend lasts 
% \LEt{***=still ongoing, or lasted? (=trend is finished)} 
for $30\Myr$. Based on our description of the SB, the first $30\Myr$ is the expanding phase of the SB, causing an additional acceleration field (see Fig.~\ref{fig:accfield_of_bubble}). We tested the significance of the SB acceleration on a fictive open cluster by generating an open cluster with mass $10^4\msun$ and radius $5\pc,$ and evaluated   the relative acceleration originating from the SB: $|a_{\rm SB}| / (|a_{\rm SB}| + |a_{\rm OC}|) = 0.18\pm0.09$. The $\pm$ describes different stars in the cluster. We conclude that the SB can provide an additional acceleration field to alter the kinematics of open clusters. However, the further acceleration does not necessarily translate to expansion, but to extended times in the outer parts of the orbits or possibly easier evaporation. 

There have been indications that the rotation of a DM halo   affects the baryonic parts of the galaxy \citep{Herpich:2015, Kataria_2022, Li_2023}, or the opposite \citep{Nu_ez_Casti_eyra_2023}. In Fig.~\ref{fig:DM_rot} we evaluate the extent to which SBs can cause the angular momentum transfer, which is about $5\%$ per $10\Gyr$. 
% \LEt{***yes?  } 
By measuring extremely precisely  how much the gas discs shrink in a galaxy absent of spirals and bars  (e.g. using chemical imprints), it is, in theory, possible to infer the rotation of the DM halo. 

\section{Summary}\label{sec:summary}
Galaxies evolve via stochastic and secular processes, for example  two-body relaxation, virialisation, momentum transfer by spirals and bars, major mergers, minor mergers. The last item in the list can provide torques via dynamical friction, changing the angular momentum of the merger with the disc. We studied additional effects related to dynamical friction that are caused not by external satellites but by gas inhomogeneities,  namely SBs introduced by multiple SN explosions originating from open clusters. 

We characterised the SB influence by acceleration field correction originating from density field correction. We applied this correction to stellar and DM particle trajectories that pass through the SB and evaluated the velocity or angular momentum changes. By stacking together all the particles, including the time-dependence of the SB, and counting the constant formation-destruction cycle of the SBs, we measured the extent of the dynamical friction from the SBs. 

Our summary of the SB influences on the galaxy is as  follows:
\begin{itemize}
    \item The acceleration field of the SB correction is directed outwards: if the gas mass is missing from the SB centre, the correction to mass is negative, and the radial acceleration is also negative (see Sect.~\ref{app:bubble_calc_formulae} for the derivation of the effect).
    \item Stars entering the SB slowly can reverse their movement direction, indicating that stars are almost co-moving with the SB (see Sect.~\ref{sec:analytical_SB}).
    \item The SB effects can be approximated empirically (see Eq.~\eqref{eq:ch_approx}). We provided an analytical approximation for a gas slab in stellar distribution that is homogeneous and isotropic. 
    \item For the test galaxy, we estimated the change to the gas disc when both the SB and the galactic potential influence the orbits of stars. The relative rate of angular momentum changes is up to $4\% \,{\rm Gyr^{-1}}$, which can account for significant changes over the galaxy's lifetime. 
    \item Although SBs are able to produce changes in galaxy disc over extended timescales, there are processes, such as an instability that drives both turbulence and inward mass migration in galaxies \citep{Goldbaum:2015, Goldbaum:2016}, that do it faster. Instabilities drive gas inwards at the rate $\dot{M}=0.1-1\,{\rm M_\odot\,yr^{-1}}$, compared to SBs $\dot{M} \lesssim 0.02\,{\rm M_\odot\,yr^{-1}}$ (see the bottom panel of Fig.~\ref{fig:slowdown_2panels}).
    \item Dynamical friction is a process that depends on the velocity of the host environment \citep{Chandrasekhar:1943}. As we are studying processes similar to dynamical friction, we estimated the possible contribution of the rotation of a DM halo to the transfer, and whether this can be used to infer the kinematics of DM. The differences are of the order of $\sim0.3\%\,{\rm Gyr^{-1}}$, or $\sim3\%$ over the time an average disc has existed. Measuring these rates is extremely difficult, but in theory the DM rotation influences the gas disc via SBs. 
    % \item We also discussed the assumptions related to the SB evolution and environement in the Discussion. \ET{I suggest to remove this point. It is not an outcome of the paper. If you want to add something then after bullet list you can say this and then add what this discussion means for the future work.}
\end{itemize}

 \begin{acknowledgements}
We thank Punyakoti Ganeshaiah Veena for helpful discussion. The present study was supported by the ETAG projects PSG700, PRG1006 and PRG2159.  This work was partially supported by the Estonian Ministry of Education and Research CoE grant "Foundations of the Universe" TK202 and the European Union's Horizon Europe research and innovation programme (EXCOSM, grant No. 101159513). AT was also supported by the COST Action CA21126 - Carbon molecular nanostructures in space (NanoSpace)-COST (European Cooperation in Science and Technology. 
 \end{acknowledgements}
\section*{Data availability}
The codes and scripts used to produce these results are available in GitLab \href{https://gitlab.ut.ee/rain.kipper/onlybubbles}{https://gitlab.ut.ee/rain.kipper/onlybubbles}. 

\bibliographystyle{aa}

\appendix

\section{Effect of dynamical friction on a spherical cavity propagating through a uniform medium}\label{app:analytic_DF}

Consider an idealised situation, in which a spherical cavity within the ISM  is propagating through a uniform ambient medium. This medium experiences no significant direct interaction with the ISM (except via gravity). For example, this cavity may result from energy deposition from multiple SNe exploding within a stellar cluster (i.e. a SB; see e.g. \citealt{Yadav:2017}), which can inflate a quasi-spherical region with a density several orders of magnitude below the ambient ISM. The ambient medium under consideration may \KT{be} associated with the DM halo of the galaxy as well as field stars unassociated with the cluster and/or bubble. 
If the relative velocity between the bubble and the ambient medium is non-zero on average (this notion will be made more precise below), the particles comprising this medium that propagate through the bubble exert a non-zero net force on the bubble (more precisely, the bubble walls) due to dynamical friction.

First, let us consider a single particle of mass $m$ propagating through a spherical cavity of radius $R$ with impact parameter $b$. According to Newton's shell theorem, the gravitational influence of a spherical density distribution can be reduced to the case where the mass internal to the test particle radius $r$ from the symmetry centre is concentrated into said centre. The theorem is also applicable to a spherically symmetric ``hole'' within an otherwise uniform medium (ISM); in this case, if $r < R$, the effective internal mass is $M(<r) = -4\pi r^3 \rho_{\rm ISM}/3$, where $\rho_{\rm ISM}$ is the ISM mass density, and vanishes if $r > R$ (here we have assumed that the material evacuated from the cavity is concentrated in a narrow region near its edge and that the total mass is conserved). Hence, while the particle is at $r > R$, it propagates within the large-scale galactic gravitational potential as if the cavity was not present, while at $r<R$ it experiences an effective extra contribution from the ''missing" mass internal to radius $r$.

\subsection{Cavity of fixed radius}

Let us first consider the case in which the radius of the cavity does not change appreciably over timescales of interest. If the mass of the particle entering the cavity satisfies $m \ll M$, then
the total energy of the particle before and after the encounter remains constant in the frame of the perturber (cavity;  i.e. $\Delta v = 0$). If the particle velocity further satisfies the condition $v^2 \gg G\rho_{\rm ISM} R^2$, its trajectory is altered only slightly upon traversing the cavity (compared with the case where the bubble is absent). Hence, the change $\Delta\vec{v}$ in particle velocity is approximately perpendicular to $\vec{v}$ after the encounter. The perpendicular component of acceleration at $r<R$ can be written as
\begin{align}
    \dot{v}_{\perp} \approx \frac{G M(<r)}{r^2} \cos{\phi} = \frac{4\pi G \rho_{\rm ISM} b}{3},
\end{align}
where
$\phi$ is the angle between the pericentre and the current position of the particle (measured from the centre of the cavity), and $b$ is the impact parameter. The total change $\Delta\vec{v}_{\perp}$ is obtained by integrating over the time interval that the test particle spends within the cavity. Writing $dt = d(b\tan{\phi}/v) \approx b d\phi/\cos^2\phi$ and observing that $r<R$ requires $-\sqrt{R^2 - b^2}/b < \tan \phi < \sqrt{R^2 - b^2}/b$, one obtains
\begin{align}
\Delta\vec{v}_{\perp} &= \int_{\phi_{\rm min}}^{\phi_{\rm max}} \frac{4\pi G \rho_{\rm ISM} b}{3} \frac{b d\phi}{v\cos^2\phi}
= \left. \frac{4\pi G \rho_{\rm ISM} b^2}{3 v} \tan \phi \right|_{-\sqrt{R^2 - b^2}/b}^{\sqrt{R^2 - b^2}/b} \nonumber \\
&= \frac{8\pi G \rho_{\rm ISM}}{3 v} b \sqrt{R^2 - b^2}.
\end{align}

Since the scalar $v$ remains unchanged by the encounter, the parallel component of the velocity change can be approximated as
\begin{align}
\Delta v_{\parallel} = \sqrt{v^2 - v_{\perp}^2} - v \approx -\frac{v}{2} \left(\frac{\Delta v_{\perp}}{v}\right)^2 = -\frac{32\pi^2 G^2\rho_{\rm ISM}^2}{9 v^3} b^2 \left( R^2 - b^2 \right).
\label{eq:dv_para}
\end{align}
The corresponding transfer of parallel momentum is $\Delta p_{\parallel} = m \Delta v_{\parallel}$.

The rate of momentum transfer (i.e. the effective drag force) for a collimated beam of particles of density $n$ is obtained by multiplying (\ref{eq:dv_para}) by the particle flux $nv$ and integrating over the impact parameter: 
\begin{align}
F_{\parallel, \rm coll} &= \int \Delta p_{\parallel} \,\, nv \, 2\pi b db = -\frac{64\pi^3 G^2\rho_{\rm ISM}^2}{9 v^2} \rho \int_0^R b^3 \left( R^2 - b^2 \right) \,db \nonumber \\
&= -\frac{16\pi^3 G^2 \rho_{\rm ISM}^2 \rho R^6}{27 v^2}.
\end{align}
Here $\rho = mn$ is the mass density of incident particles and $2\pi b db$ is a surface element perpendicular to the particle flux.

For an arbitrary distribution of particles, one should substitute
\begin{align}
    \rho \rightarrow m \int f(\vec{v}) d^3 v,
\end{align}
where $f(\vec{v})$ is the phase space density, to obtain
\begin{align}
    \vec{F} = -\frac{16\pi^3 G^2 \rho_{\rm ISM}^2 R^6}{27} m \int \frac{f(\vec{v})}{v^2} \vec{\omega} \, d^3 v,
    \label{eq:Fvec}
\end{align}
where $\vec{\omega} = \vec{v}/v$ is a unit vector in the direction of the incident particle velocity.

If the incident particle distribution is isotropic, then Eq. (\ref{eq:Fvec}) shows that the net momentum exchange vanishes, as is expected from symmetry considerations. Next, let us consider the simplest non-trivial case, in which the phase space density is isotropic in a frame that propagates with a velocity $\vec{v}_0$ relative to the perturber (cavity). By symmetry, the dynamical friction force must be parallel to $\vec{v}_0$, hence we  consider only the $\vec{F}\parallel \vec{v}_0$ component when taking the angular integral. Making use of the fact that $f(\vec{v})$ is invariant upon Galilean transformations (or indeed Lorentz transformations if written in terms of $\vec{p}$) and so is $d^3 v$, one obtains
\begin{align}
    F = -\frac{16\pi^3 G^2 \rho_{\rm ISM}^2 R^6}{27} m \int \frac{f^{\prime}(v^{\prime})}{v^2} \cos\theta \, d^3 v^{\prime},
    \label{eq:Fvec2}
\end{align}
where $\vec{v}^{\prime} = \vec{v} - \vec{v}_0$ is the particle velocity in the isotropy frame and
$\cos\theta = \omega\cdot \vec{v}_0/v_0 = \vec{v} \cdot \vec{v}_0/(v v_0)$. Using the cosine rule to express $\cos\theta = (v^2 + v_0^2 - v^{\prime \, 2})/2vv_0$ and $v^2 = v^{\prime \, 2} + v_0^2 + 2 v^{\prime}v_0 \cos\theta^{\prime}$, where $\cos\theta^{\prime} = \vec{v^{\prime}} \cdot \vec{v}_0/(v^{\prime} v_0)$, one can write Eq. (\ref{eq:Fvec2}) as
\begin{align}
    F &= -\frac{16\pi^3 G^2 \rho_{\rm ISM}^2 R^6}{27} m \int f^{\prime}(v^{\prime}) \, v^{\prime \, 2} dv^{\prime}
    \nonumber \\
    &\times \int \frac{1 + q \cos\theta^{\prime}}{v_0^2 (1 + q^2 + 2q \cos\theta^{\prime})^{3/2}} \, d\phi^{\prime} d\cos\theta^{\prime}.
    \label{eq:Fvec3}
\end{align}
Here we have expressed the velocity space element as $d^3 v^{\prime} = v^{\prime \, 2} dv^{\prime} d\phi^{\prime} d\cos\theta^{\prime}$, where $d\phi^{\prime} d\cos\theta^{\prime}$ is the solid angle in the isotropy frame, and defined $q \equiv v^{\prime}/v_0$. The integral over the azimuthal angle $\phi^{\prime}$ is trivial and contributes a factor $2\pi$. The integral over $\cos\theta$ is also elementary, yielding $2/v_0^2$, hence the final result is
\begin{align}
    F &= -\frac{64\pi^3 G^2 \rho_{\rm ISM}^2 R^6 \,m}{27 v_0^2} \int f^{\prime}(v^{\prime}) \, v^{\prime \, 2} dv^{\prime}.
    \label{eq:Fvec4}
\end{align}

\subsection{Cavity of time-dependent radius}
\label{app:SB:timedep}

In the case of an expanding or contracting cavity the entry and exit radii of a particle traversing it will be different and the kinetic energy of the latter is no longer conserved by the encounter even if $m \ll M$. 

In the frame of the cavity, the radial component of a particle's equation of motion within the cavity reads
\begin{align}
\frac{dv_r}{dt} = \frac{d}{dt} \left( \vec{v}\cdot\frac{\vec{r}}{r} \right) =
F_r + \frac{L^2}{r^3},
\label{eq:app:dvdr}
\end{align}
where $L = r^2 \dot{\theta}$ is the (conserved) angular momentum of the particle.
Using the relation $dv_r/dt = (1/2) dv_r^2/dr$ and writing $F_r = -d\Phi(r)/dr$, one obtains
\begin{align}
\frac{d}{dr}\left( \frac{v_r^2}{2}  + \Phi + \frac{L^2}{2r^2} \right) = 0,
\label{eq:app:E}
\end{align}
or simply $v^2/2 + \Phi = E = \mbox{constant}$, where $E$ is the conserved total energy.

For the cavity we can write
\begin{align}
\Phi(r) = -\frac{2\pi G \rho_{\rm ISM}}{3} r^2 = -\frac{1}{4} v_{\rm ch}^2 \frac{r^2}{R_{\rm enter}^2},
\end{align}
where $R_{\rm enter}$ is the entry radius and
we have introduced a characteristic velocity
\begin{align}
v_{\rm ch} = \sqrt{\frac{8\pi G \rho_{\rm ISM} R_{\rm enter}^2}{3}}.
\label{eq:app:vff}
\end{align}
Applying the energy conservation condition, one obtains the kinetic energy gain as a function of entry and exit radii into/from the bubble,
\begin{align}
v_{\rm out}^2 = v_{\rm in}^2 + \frac{v_{\rm ch}^2}{2} \left[\left(\frac{R_{\rm exit}}{R_{\rm enter}}\right)^2 - 1\right].
\label{eq:app:vout}
\end{align}

For a given $R_{\rm enter}$, the exit radius $R_{\rm exit}$ can be determined by solving Eq.~(\ref{eq:app:E}) for $r(t)$ and subsequently eliminating $t$ from $r(t) = r_{\rm exp}(t)$, where $r_{\rm exp}(t)$ characterises the evolution of the bubble radius. In general, the latter step has to be performed numerically. For a spherically symmetric bubble, $R_{\rm exit}$ is a function of entry time $t_{\rm in}$ (which also determines $R_{\rm enter} = r_{\rm exp}(t_{\rm in})$), particle velocity $v$, and the angle $\theta$ between particle velocity and the local bubble surface normal at $t_{\rm in}$.

\subsubsection{Integration over phase space}
\label{app:SB:PhSp}

Once the energy/momentum gain or loss of a single particle entering the bubble at a given $t$ and $R_{\rm enter} = r_{\rm b}(t)$ is known, the next task is to determine the net transfer rate of a population of particles characterised by a phase space density $f(\vec{v})$. For this we first need to find the number of particles crossing the bubble surface element in unit time. The inward particle flux in the frame comoving with the surface element is (defined as positive definite)
\begin{align}
dF_{\rm in} = -\left(\vec{v}^{\dagger}\cdot \frac{\vec{v}_{\rm exp}}{v_{\rm exp}} \right) f^{\dagger}(\vec{v}^{\dagger}) d^3 v^{\dagger},
\label{eq:app:dF}
\end{align}
where $\vec{v}^{\dagger} = \vec{v} - \vec{v}_{\rm exp}$ is the particle velocity in the surface element frame, $\vec{v}$ is its velocity in the bubble frame (i.e. its centre), and $\vec{v}_{\rm exp}$ is the local velocity of the surface element in the same frame. The scalar product in Eq.~(\ref{eq:app:dF}) can be written as $-\vec{v}^{\dagger}\cdot \vec{v}_{\rm exp}/v_{\rm exp} = -|v|\cos\theta + v_{\rm exp} > 0$, where $\cos\theta$ is the angle between the particle velocity and the local radial direction in the bubble frame. We note that this quantity has to be positive, which places a constraint on the angles at which particles of given a velocity can enter the bubble. The particle flux thus becomes
\begin{align}
dF_{\rm in} =  (v_{\rm exp} -|v|\cos\theta) \, f(\vec{v}) d^3 v,
\quad\mbox{where} \quad
\cos\theta < \frac{v_{\rm exp}}{|v|},
\label{eq:app:dF2}
\end{align}
and we have again used the invariance of $f$ and $d^3 v$ by dropping the symbol $\dagger$. The normalisation of $f$ gives the number density of stars, or physical density if multiplied with the mass of a star
\begin{align}
    m\int f(\vec{v}){\rm d}v = \rho^* \label{eq:f_normalisation}
\end{align}

Next, let us assume that a reference frame exists in which the particle distribution is isotropic (referred to below as the lab frame), and that the bubble propagates with a velocity $\vec{v}_{\rm b}$ relative to this frame.
The total rate of momentum transfer between the bubble and the particle population can then be written as
\begin{align}
\frac{dP}{dt} = \int m\Delta v_{\parallel} \cos\theta_v \, dF_{\rm in} \, dS_{\rm b}.
\label{eq:app:dPdt}
\end{align}
Here $m\Delta v_{\parallel}$ is the
transferred momentum of a single particle parallel to its initial direction of propagation (the net transfer of transverse momentum vanishes by symmetry in a homogeneous particle field and a spherically symmetric perturber), $\theta_v$ is the angle between the velocity $\vec{v}$ of the incoming particle and the bubble velocity $\vec{v}_{\rm b}$ measured in the bubble frame, and $dF_{\rm in}$ is the particle flux defined by Eq.~(\ref{eq:app:dF2}).
The bubble surface element is 
\begin{align}
    dS_{\rm b} = R_{\rm enter}^2 d\cos\theta \, d\phi,
    \label{eq:app:dS}
\end{align}
where the angles refer to the polar and azimuthal angles of the local surface normal vector in a coordinate system anchored to the incoming particle velocity $\vec{v}$ (in the bubble frame). This choice is motivated by the fact that in this coordinate system the integrand of Eq.~(\ref{eq:app:dPdt}) is independent of $\phi$ and the corresponding integral thus immediately yields $2\pi$.

To make further use of the symmetries in the problem, the integral over particle velocities is best taken in the frame where the distribution is isotropic (i.e. lab frame, denoted by prime) by writing $f(\vec{v}) d^3 v = f^{\prime}(\vec{v}^{\prime}) d^3 v^{\prime} = f^{\prime}(v^{\prime}) \, v^{\prime \, 2} dv^{\prime} \, d\cos\theta_v^{\prime} \, d\phi_v^{\prime}$, where $\theta_v^{\prime}$ denotes the angle between $\vec{v}^{\prime}$ and $\vec{v_{\rm b}}$. By symmetry, the integral over the corresponding azimuth $\phi_v^{\prime}$ is again trivial and gives $2\pi$.

To proceed further, we need the relations between particle velocities and angles in the lab and bubble frames. Defining $\beta_{\rm b}^{\prime} = v_{\rm b}/v^{\prime}$, one obtains
\begin{align}
    \cos\theta_v = \frac{\cos\theta_v^{\prime} - \beta_{\rm b}^{\prime}}{\sqrt{1 - 2\cos\theta_v^{\prime}\beta_{\rm b}^{\prime} + \beta_{\rm b}^{\prime \, 2}}}, \quad
    v = v^{\prime} \sqrt{1 - 2\cos\theta_v^{\prime}\beta_{\rm b}^{\prime} + \beta_{\rm b}^{\prime \, 2}}.
\end{align}
%Using the above relations,
Putting all of the above together,
Eq.~(\ref{eq:app:dPdt}) takes the form
\begin{align}
\frac{dP}{dt} &= 4\pi^2 m R_{\rm enter}^2 \int_{-\infty}^{\infty} f^{\prime}(v^{\prime}) \, v^{\prime \, 3} dv^{\prime}
\int_{-1}^1 (\cos\theta_v^{\prime} - \beta_{\rm b}^{\prime}) \, d\cos\theta_v^{\prime} \nonumber \\
&\times \int_{D_{\theta}} (\beta - \cos\theta) \, \Delta v_{\parallel} \, d\cos\theta,
\label{eq:app:dPdt2}
\end{align}
where
\begin{align}
    D_{\theta} &= \{\cos\theta : \cos\theta \ge -1 \land \cos\theta \le \min(1,\beta)\}
\end{align}
and $\beta \equiv v_{\rm exp}/|v|$.

\section{SB collapse analysis}\label{app:SB_diffusion}
One of the shortcomings of the current analysis is the unclear behaviour of a SB after the SNe cease depositing energy and momentum into the SB (see Sect.~\ref{sec:discussion_SB_assumptions}). Here we are testing how sensitive our results are to the specific form of the evolution of the SB after this cease, at $t_{\rm max}$ (Sect.~\ref{sec:bubble_evolution} explains the fiducial behaviour). 
We are analysing three ways how the functional form of SB can be modified: separation of inner and outer edge collapsing speeds, then the vanishing of the SB by not changing radii in the collapsing phase, but the density amplitude; and last the influence of shearing to collapse. These modifications somewhat mimic different processes that may happen, we  analyse them in the given order. 

% There are three relatively simple ways to modify the SB evolution: changing the collapsing of inner radius, changing the collapse of outer radius, and drifting from collapsing model to altering only densities. We do not use the latter approach as it would violate the momentum and continuity equations (gas needs to travel from the shell parts of the SB to the centre very fast, which does not seem physical). }

\begin{figure}
    \centering
    \includegraphics{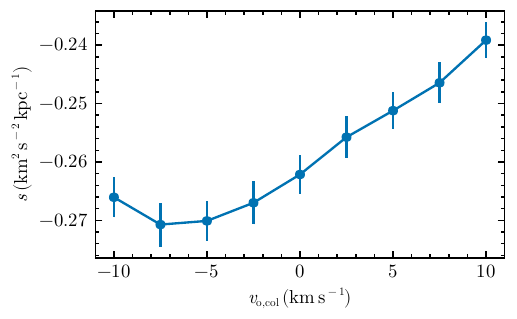}
    \caption{The dependence of de-acceleration on the outer shell $R_{\rm o}$ collapse speed ($v_{\rm o, col}$, the negative value indicates continuous expansion, the positive collapse of radius). While keeping the inner radius collapse speed ($v_{\rm col} = 10\,{\rm km\,s^{-1}}$) as constant. We can see that the de-acceleration remains relatively constant throughout various speeds, indicating that the outer shell collapse speed alters the overall transfer of gas angular momentum only little. }
    \label{fig:v_outer_collapse}
\end{figure}

By relaxing our treatment of SB collapsing speeds of inner and outer radii ($R_{\rm i}$ and $R_{\rm o}$), it allows us to mimic the outer shell stalling that is predicted for atmospheric conditions \citep{Penner:2003}. In the main part of the paper we had one constant to describe both ($v_{\rm col}$), but in the current analysis we split it into two separate speeds: inner (still $v_{\rm col}$) and outer ($v_{\rm o, col}$) shell. To understand how these changes may influence we remind that $R_{\rm i}$ determines what is the mass that is displaced to form the acceleration field correction and $R_{\rm o}$ effectively determines what is the shape of the acceleration field. When we keep $v_{\rm col}$ as constant, we can see how sensitive our results are to the precise acceleration field shape. The analysis of the set-up is same as in Sect.~\ref{sec:bubble_like_Chandrasekhar} -- a homogeneous environment\footnote{Including a constant value to the rate of the SBs $f_{\rm b}$}. 
Figure~\ref{fig:v_outer_collapse} shows what is the introduced de-acceleration changes when varying $v_{\rm o,col}$. The varying range is from collapsing along with inner radius to ones retaining expansion speed outwards, even after the SNe cease supporting high temperature. The de-acceleration range is $-0.27\dots-0.24$, or about $\pm\approx6\%$. We conclude that the behaviour of outer shell does not alter de-acceleration ($s$) much. 

\begin{figure}
    \centering
    \includegraphics{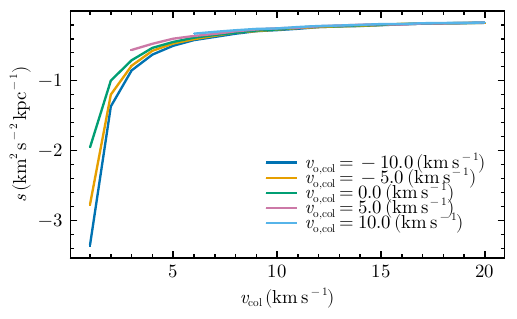}
    \caption{The dependence of the de-acceleration on the inner radius ($R_{\rm i}$) collapsing speeds ($v_{\rm col}$). In the calculation, the inner radius reaches $0$ at different times, hence the lifetime of SBs are different and overall contribution per SB is elevated. 
    % For reference, both the inner and outer radius collapsed with the speed of $10\,{\rm km\,s^{-1}}$ in the main text analysis. 
    In the analysis we kept the rate of SB formation ($f_{\rm b}$) as constant. The lines show different outer radius collapsing speeds, and some do not reach small values as the outer radius of SB would shrink below inner radius, which we consider physically inconsistent.  }
    \label{fig:v_col_inner_dep}
\end{figure}

The second aspect of the first test is aimed to see the sensitivity of $R_{\rm i}$ collapse speed. We kept the $R_{\rm o,col}$ as constant and calculated the de-acceleration dependence on the $v_{\rm col}$. The corresponding result is shown in Fig.~\ref{fig:v_col_inner_dep} -- they are qualitatively same as in Fig.~\ref{fig:slowdown_2panels}: very slow collapses can provide about an order of magnitude larger contributions than the fast ones. There is little differences in faster outer shell collapse speeds. Overall, we conclude that the specific form of collapsing phase does not alter the results much. 
% \textbf{Our first test involves the sensitivity on the evolution of $R_{\rm o}$. Here, we modified Eq.~\eqref{eq:Ro_fid} so that the collapsing speed of the outer part of the SB ($v_{\rm o, col}$ from here and onward) is independent of the inner one (still $v_{\rm col}$). The setup for analysis is the same as in Sect.~\ref{sec:bubble_like_Chandrasekhar} -- a homogeneous environment\footnote{Including a constant value to the rate of the SBs $f_{\rm b}$. It is unclear if this should be constant as we assume constant SF, that causes constant rate of SBs. As an opposing argument, the disc can accommodate only limited number of SBs without the shapes of SBs being distorted, which may be overthrown when the lifetime of SB increases too much. }. The de-acceleration dependence on the $v_{\rm o,col}$ is shown in Fig.~\ref{fig:v_outer_collapse}, from where we can conclude that the variations of the collapsing speed can produce de-acceleration differences within the range $-0.27\dots-0.24$, or about $\pm\approx6\%$ with very different input range (from collapsing to keeping expanding). }
% \textbf{Our second test involves learning the sensitivity to the changes of inner radius collapse while the outer radius is kept constant. As displacement of the mass in the inner radius determines the displaced mass, hence the amplitude of the  acceleration, we suspect it to have larger influence. We provide the numerical results in Fig.~\ref{fig:v_col_inner_dep}. }
% \textbf{The results are qualitatively same behaviour as Fig.~\ref{fig:slowdown_2panels}: very slow collapses can provide about order of magnitude larger contributions than fast ones. Overall, we conclude that the specific form of collapsing phase does not alter the results much. }

\begin{figure}
    \centering
    \includegraphics[width=1\linewidth]{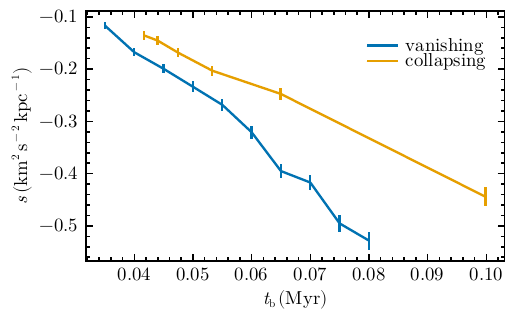}
    \caption{The dependence of de-acceleration on the type of collapsing. Vanishing describes the situation where the SB density is altered according to Eq.~\ref{eq:acc_vanish_version} and collapsing to the one where $v_{\rm col} = 10\,{\rm km\,s^{-1}}$ and $v_{\rm o,col} = 0\,{\rm km\,s^{-1}}$.}
    \label{fig:vanish_deacc}
\end{figure}

Another scenario for how the SB may collapse is by intrusion of matter into the SB region, for example turbulence eddies or instability imprints entering the SB. These are not symmetric entities, hence they cannot be described in a symmetric manner, and we take a statistical approach: we re-scale the acceleration field after $t_{\rm max}$ with a variable that is $1$ at $t_{\rm max}$ and will drop linearly to $0$ upon the end of SB life $t_{\rm b}$. So between $t_{\rm max}\dots t_{\rm b}$ we have acceleration of SB as
\begin{equation}
    {\bf a} = {\bf a}^{\rm fid}({\rm t = t_{\rm max}}) \frac{t_{\rm b}  - t}{t_{\rm b} - t_{\rm max}}. \label{eq:acc_vanish_version}
\end{equation}
The outcome is shown in Fig.~\ref{fig:vanish_deacc}. Since the collapsing and the vanishing versions are not directly comparable we chose the lifetime of SB as depending variable. We can see that the vanishing description of the SB has a stronger impact on the de-acceleration, hence our results are more conservative, or underestimated. The physical reason why the vanishing SB influences  de-acceleration more strongly than the collapsing model does is the amount of mass that is displaced. In the case of the collapsing model it shrinks with radius in cubic form when the radius decreases linearly, but in the case of the vanishing model, it is a linear decrease. 

\begin{figure}
    \centering
    \includegraphics[width=1.0\linewidth]{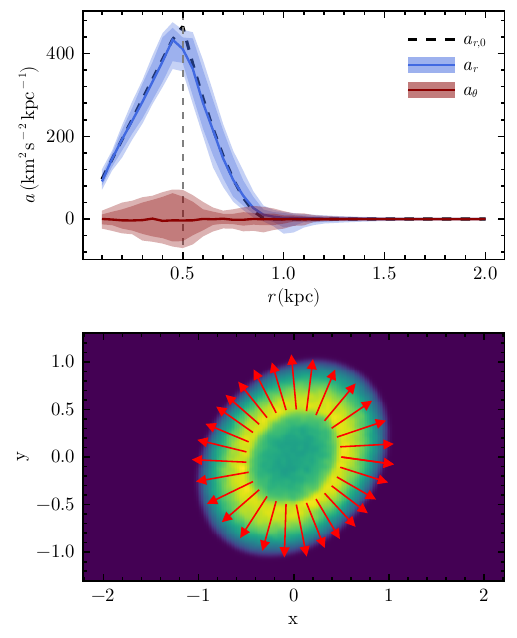}
    \caption{Acceleration field of sheared SB. The top panel shows   the acceleration field components of a close-to-maximally sheared SB mass distribution compared to a non-shared distribution. The corridors describe the $65$ and $95\%$ range of the components with different directions from the centre. The dashed line shows at what radius the bottom panel is measured.  In the bottom panel we depict how the acceleration field changes with direction from the centre of the SB. The tangential component and the angular dependence of acceleration field can be observed.}  
    \label{fig:shear_max}
\end{figure}

The last possibility that we analyse for the SB collapse is the shearing of the SB due to differential rotation of galaxies. The amount of shearing depends on the SB location in the galaxy: inner regions have higher angular speeds than outer regions, and therefore shear the  SB more. Since our used method breaks down upon distortions from sphericity (spherical symmetry of density was needed to evaluate SB acceleration field and to keep the stellar orbit integration bounds constrained to finite region) we are limited for only testing. The sheared SB is constructed by sampling particles according to the total gas density profile (both uniform and SB-containing), and shearing their location linearly according to the range required for the test. Hence we relied on evaluating the acceleration field of the sheared SN and compared it with non-sheared one. The chosen amount of shearing was chosen based on observations: \citet{Watkins:2023} provided the SB axis lengths for NGC 628, we used their catalogue of SBs to select only SBs larger than $100\pc$ and found that the minimum axis ratio was $0.56$, and mean $0.75$. We sheared the SB to the value close to the extreme of observed ones: $0.6$. The resulting acceleration field is shown in Fig.~\ref{fig:shear_max}. We concluded from the figure that maximally sheared SB can cause $6-39\%$ offsets from homogeneous case at $r = 0.5\kpc$. The same numbers for SB with axis ratio of $0.75$ provide $2-23\%$ offsets. % 0.77, 0.98$

\label{lastpage}
\end{document}